\newcommand\electron{$\mathrm{e}^{-}$}
\newcommand\positron{$\mathrm{e}^{+}$}
\newcommand\pbar{$\bar{\mathrm{p}}$}
\newcommand\antihydrogen{$\bar{\mathrm{H}}$}
\newcommand\alphag{\mbox{ALPHA-\textscriptg}}
\newcommand\alphatwo{\mbox{ALPHA-2}}

\documentclass[reprint, superscriptaddress]{revtex4-2}

\usepackage{amsmath, amssymb, gensymb, tipa}

\usepackage{graphicx, float}
\graphicspath{{Figures/}}

\usepackage[unicode, colorlinks=true, linkcolor=black, urlcolor=blue, citecolor=blue]{hyperref}

\usepackage{xcolor}

\begin{document}
\title{\textbf{Design and Performance of A Novel Low Energy Multi-Species \\ Beamline for the ALPHA Antihydrogen Experiment}}

\author{C. J. Baker}
\affiliation{Department of Physics, College of Science, Swansea University, Swansea SA2 8PP, UK}

\author{W. Bertsche}
\affiliation{School of Physics and Astronomy, University of Manchester, Manchester M12 9PL, UK}
\affiliation{Cockcroft Institute, Sci-Tech Daresbury, Warrington WA4 4AD, UK}

\author{A. Capra}
\affiliation{TRIUMF, 4004 Wesbrook Mall, Vancouver, BC, Canada V6T 2A3}

\author{C.L. Cesar}
\affiliation{Instituto de Fisica, Universidade Federal do Rio de Janeiro, Rio de Janeiro 21941-972, Brazil}

\author{M. Charlton}
\affiliation{Department of Physics, College of Science, Swansea University, Swansea SA2 8PP, UK}

\author{A.J. Christensen}
\affiliation{Department of Physics, University of California at Berkeley, Berkeley, CA 94720-7300, USA}

\author{R. Collister}
\affiliation{TRIUMF, 4004 Wesbrook Mall, Vancouver, BC, Canada V6T 2A3}

\author{A. Cridland Mathad}
\affiliation{Department of Physics, College of Science, Swansea University, Swansea SA2 8PP, UK}

\author{S. Eriksson}
\affiliation{Department of Physics, College of Science, Swansea University, Swansea SA2 8PP, UK}

\author{A. Evans}
\affiliation{Department of Physics and Astronomy, University of Calgary, Calgary AB, Canada T2N 1N4}

\author{N. Evetts}
\affiliation{Department of Physics and Astronomy, University of British Columbia, Vancouver BC, Canada V6T 1Z1}

\author{S. Fabbri}
\affiliation{School of Physics and Astronomy, University of Manchester, Manchester M12 9PL, UK}

\author{J. Fajans}
\affiliation{Department of Physics, University of California at Berkeley, Berkeley, CA 94720-7300, USA}

\author{T. Friesen}
\affiliation{Department of Physics and Astronomy, University of Calgary, Calgary AB, Canada T2N 1N4}

\author{M.C. Fujiwara}
\affiliation{TRIUMF, 4004 Wesbrook Mall, Vancouver, BC, Canada V6T 2A3}

\author{D.R. Gill}
\affiliation{TRIUMF, 4004 Wesbrook Mall, Vancouver, BC, Canada V6T 2A3}

\author{P. Grandemange}
\affiliation{TRIUMF, 4004 Wesbrook Mall, Vancouver, BC, Canada V6T 2A3}

\author{P. Granum}
\affiliation{Department of Physics and Astronomy, Aarhus University, DK-8000 Aarhus C, Denmark}

\author{J.S. Hangst}
\affiliation{Department of Physics and Astronomy, Aarhus University, DK-8000 Aarhus C, Denmark}

\author{M.E. Hayden}
\affiliation{Department of Physics, Simon Fraser University, Burnaby BC, Canada V5A 1S6}

\author{D. Hodgkinson}
\affiliation{School of Physics and Astronomy, University of Manchester, Manchester M12 9PL, UK}

\author{C.A. Isaac}
\affiliation{Department of Physics, College of Science, Swansea University, Swansea SA2 8PP, UK}

\author{M. A. Johnson}
\affiliation{School of Physics and Astronomy, University of Manchester, Manchester M12 9PL, UK}
\affiliation{Cockcroft Institute, Sci-Tech Daresbury, Warrington WA4 4AD, UK}

\author{J.M. Jones}
\affiliation{Department of Physics, College of Science, Swansea University, Swansea SA2 8PP, UK}

\author{S.A. Jones}
\affiliation{Department of Physics and Astronomy, Aarhus University, DK-8000 Aarhus C, Denmark}

\author{A. Khramov}
\affiliation{TRIUMF, 4004 Wesbrook Mall, Vancouver, BC, Canada V6T 2A3}

\author{L. Kurchaninov}
\affiliation{TRIUMF, 4004 Wesbrook Mall, Vancouver, BC, Canada V6T 2A3}

\author{N. Madsen}
\affiliation{Department of Physics, College of Science, Swansea University, Swansea SA2 8PP, UK}

\author{D. Maxwell}
\affiliation{Department of Physics, College of Science, Swansea University, Swansea SA2 8PP, UK}

\author{J.T.K. McKenna}
\affiliation{Department of Physics and Astronomy, Aarhus University, DK-8000 Aarhus C, Denmark}

\author{S. Menary}
\affiliation{Department of Physics and Astronomy, York University, Toronto, ON M3J 1P3, Canada}

\author{T. Momose}
\affiliation{Department of Physics and Astronomy, University of British Columbia, Vancouver BC, Canada V6T 1Z1}

\author{P.S. Mullan}
\affiliation{Department of Physics, College of Science, Swansea University, Swansea SA2 8PP, UK}

\author{J.J. Munich}
\affiliation{Department of Physics, Simon Fraser University, Burnaby BC, Canada V5A 1S6}

\author{K. Olchanski}
\affiliation{TRIUMF, 4004 Wesbrook Mall, Vancouver, BC, Canada V6T 2A3}

\author{J. Peszka}
\affiliation{Department of Physics, College of Science, Swansea University, Swansea SA2 8PP, UK}

\author{A. Powell}
\affiliation{Department of Physics and Astronomy, University of Calgary, Calgary AB, Canada T2N 1N4}

\author{C.\O. Rasmussen}
\affiliation{Experimental Physics Department, CERN, Geneva 1211, Switzerland}

\author{R.L. Sacramento}
\affiliation{Instituto de Fisica, Universidade Federal do Rio de Janeiro, Rio de Janeiro 21941-972, Brazil}

\author{M. Sameed}
\affiliation{School of Physics and Astronomy, University of Manchester, Manchester M12 9PL, UK}

\author{E. Sarid}
\affiliation{Soreq NRC, Yavne, 81800, Israel}

\author{D.M. Silveira}
\affiliation{Instituto de Fisica, Universidade Federal do Rio de Janeiro, Rio de Janeiro 21941-972, Brazil}

\author{C. So}
\affiliation{Department of Physics and Astronomy, University of Calgary, Calgary AB, Canada T2N 1N4}

\author{D.M. Starko}
\affiliation{Department of Physics and Astronomy, York University, Toronto, ON M3J 1P3, Canada}

\author{G. Stutter}
\affiliation{Department of Physics and Astronomy, Aarhus University, DK-8000 Aarhus C, Denmark}

\author{T.D. Tharp}
\affiliation{Physics Department, Marquette University, P.O. Box 1881, Milwaukee, WI 53201-1881, USA}

\author{R.I. Thompson}
\affiliation{Department of Physics and Astronomy, University of Calgary, Calgary AB, Canada T2N 1N4}

\author{C. Torkzaban}
\affiliation{Department of Physics, University of California at Berkeley, Berkeley, CA 94720-7300, USA}

\author{D.P. van der Werf}
\affiliation{Department of Physics, College of Science, Swansea University, Swansea SA2 8PP, UK}

\author{J.S. Wurtele}
\affiliation{Department of Physics, University of California at Berkeley, Berkeley, CA 94720-7300, USA}

\collaboration{The ALPHA Collaboration}

\begin{abstract}
\vspace{15pt}
The ALPHA Collaboration, based at the CERN Antiproton Decelerator, has recently implemented a novel beamline for low-energy ($\lesssim$ 100 eV) positron and antiproton transport between cylindrical Penning traps that have strong axial magnetic fields.
Here, we describe how a combination of semi-analytical and numerical calculations were used to optimise the layout and design of this beamline.
Using experimental measurements taken during the initial commissioning of the instrument, we evaluate its performance and validate the models used for its development.
By combining data from a range of sources, we show that the beamline has a high transfer efficiency, and estimate that the percentage of particles captured in the experiments from each bunch is \mbox{$\left( 78 \pm 3 \right) \%$} for up to $10^{5}$ antiprotons, and \mbox{$\left( 71 \pm 5 \right) \%$} for bunches of up to $10^{7}$ positrons.
\end{abstract}

\maketitle

\section{INTRODUCTION} \label{sec:introduction}
\begin{figure*}
\renewcommand{\abovecaptionskip}{10pt}
\renewcommand{\belowcaptionskip}{0pt}
\centering{
\vspace{5pt}
\includegraphics[scale=0.33, trim=0 5 0 0, clip]{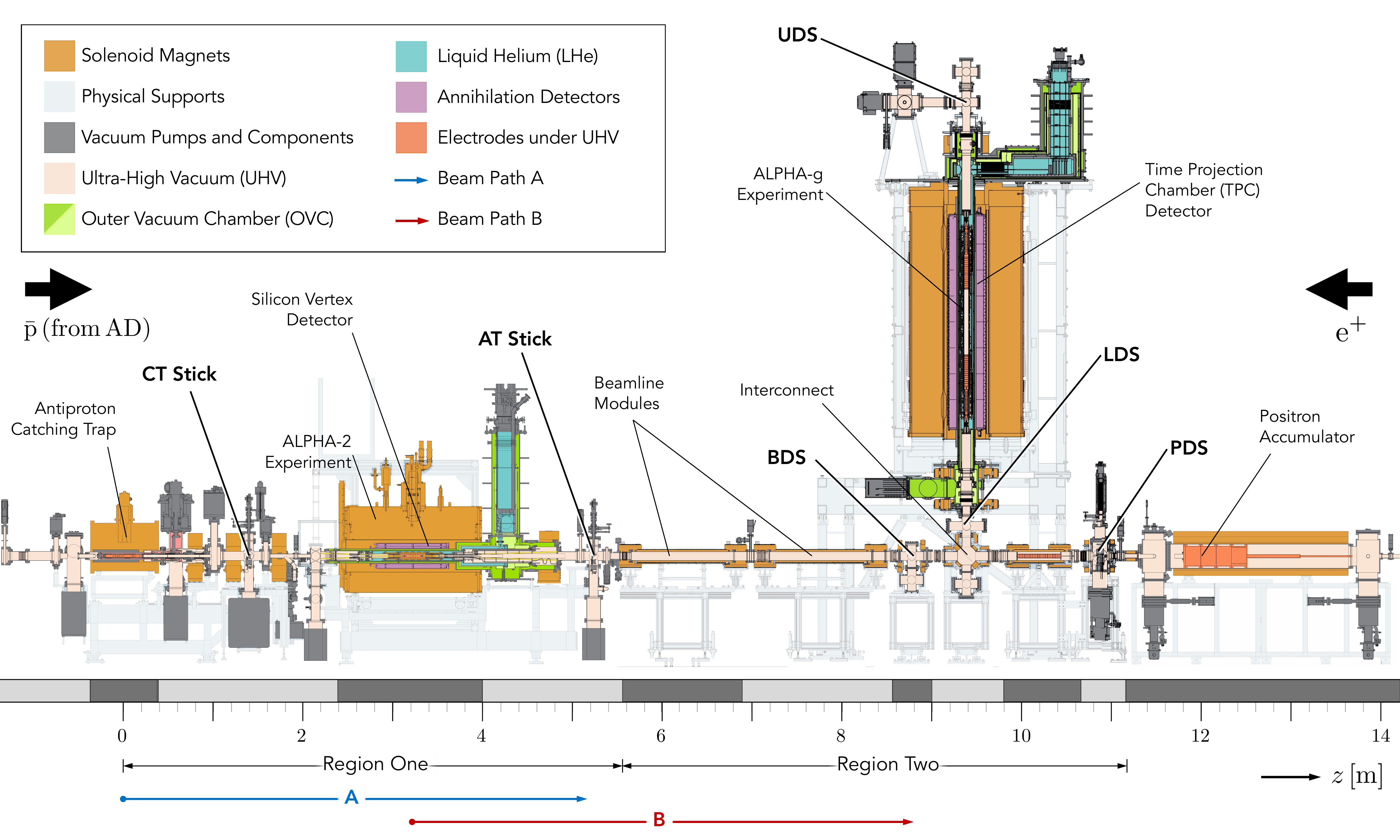}
}
\caption{Schematic showing a cross-section view of the ALPHA apparatus following the installation of the \alphag{} experiment and new beamline in 2018. The horizontal axis is shaded to differentiate between individual sections of the apparatus. Smaller detectors (e.g. scintillator panels) are not shown for clarity. The locations of beam diagnostics are annotated in bold. See text for further details. }
\label{fig:alpha-schematic}
\end{figure*}
The ALPHA experiment \cite{ALPHA-2014-1} at the CERN Antiproton Decelerator (AD) (see e.g. \cite{Hori-2013}) studies magnetically trapped antihydrogen (\antihydrogen{}) atoms, produced by merging clouds of cold positrons (\positron{}) and antiprotons (\pbar{}) \cite{ALPHA-2017-2}.
Precision measurements of trapped antihydrogen atoms provide unique, high-resolution tests of fundamental symmetries, and may help to explain why antimatter is so scarce in our universe.
In recent years, the ALPHA collaboration has succeeded in measuring several features in the antihydrogen spectrum, including its narrow 1S-2S transition \cite{ALPHA-2016-2, ALPHA-2018-1}, 1S-2P Lyman-alpha transitions \cite{ALPHA-2018-2} and ground state hyperfine splitting \cite{ALPHA-2017-1}.
These measurements already represent unprecedented tests of new physics beyond the standard model and CPT (Charge conjugation, Parity inversion and Time reversal) invariance \cite{Bluhm-1999, Charlton-2020}.
\par
In 2018, the ALPHA apparatus \mbox{(Fig. \ref{fig:alpha-schematic})} was significantly expanded with the addition of \alphag{}, a secondary, vertical atom trap intended to make direct measurements of antimatter's gravitational acceleration \cite{Bertsche-2018}.
This device operates alongside and shares much of its design with the original ALPHA apparatus \cite{ALPHA-2014-1}, employing an energetically shallow (\mbox{$\sim$ 0.54 K} in temperature units) magnetic minimum trap to confine antihydrogen atoms.
In a typical ALPHA experimental cycle, around \mbox{$10^{5}$} antiprotons are allowed to mix with a cold \mbox{($\sim$ 20 K)} plasma of $3 \times 10^{6}$ positrons by manipulating the electric potential along the axis of a cylindrical Penning trap.
Each mixing cycle yields around 20 trapped \antihydrogen{} that can be stacked for many hours and subsequently used for experiments \cite{ALPHA-2017-2}.
\par
Prior to antihydrogen production, positrons and antiprotons must be transferred from their respective source traps into one of ALPHA's two \antihydrogen{} synthesis traps.
With the installation of the \alphag{} experiment in 2018, a new charged particle beamline was required to transport \pbar{} and \positron{} clouds between the various particle traps.
However, the magnetised beams that are extracted from Penning traps (trap-based beams) have a number of properties that make the design of this beamline challenging.
\par
For practical reasons, \pbar{} and \positron{} bunches are only extracted into the ALPHA beamline with very low energies of \mbox{$\lesssim$ 100 eV}.
The use of large, unshielded superconducting magnets at ALPHA therefore rules out beam transport using conventional magnetic or electrostatic lattice beamlines, since these magnets generate stray fields of  hundreds of Gauss between sections of the apparatus.
By comparison, the magnetic field required to steer a \mbox{50 eV} \positron{} beam about a typical radius of \mbox{200 mm} is only \mbox{$\sim$ 1 Gauss} \mbox{($10^{-4}$ T)}.
\par
In addition, trap-based beams conserve a canonical angular momentum that couples their transverse size to the inverse of the local magnetic field strength (see \mbox{Sec. \ref{sec:semi-analytical}}) \cite{Danielson-2015}.
The extraction of low-energy \pbar{} and \positron{} bunches into a beamline with no residual magnetic field is therefore non-trivial \cite{Hurst-2015, Weber-2011}, generally resulting in particle losses and increased beam emittances.
\par
As a result, particles are transported using a magnetically guided beamline, where \pbar{} and \positron{} bunches are channelled through a series of solenoids that provide continual steering and focusing in the transverse plane.
This scheme avoids the need for excessively weak electromagnetic fields, and in some cases exploits the stray fields between particle traps to guide bunches along the beamline (see \mbox{Sec. \ref{sec:region-one}}).
Furthermore, this type of transport does not require the extraction of particles into zero magnetic field.
\par
A range of semi-analytical and numerical models were used to develop the design of the ALPHA beamline.
In many cases, the beam dynamics can be approximated using simple models such as the Guiding Centre Approximations (GCA) \cite{Chen-1984}.
However, in regions where the magnetic field is very weak or inhomogeneous, individual \pbar{} and \positron{} particles can adopt more complex motions that are impossible to model analytically.
In this regime, numerical particle tracing simulations were required to accurately solve the equation of motion for each particle.
\par
The structure of this paper is as follows:
Section \ref{sec:sources} describes the \pbar{} and \positron{} sources in use at ALPHA as pertains to this study.
Section \ref{sec:theory} reviews semi-analytical and numerical methods that were used to model the dynamics of charged particle bunches transported along the new beamline.
Section \ref{sec:design} provides an overview of the instrument design that was implemented at CERN during 2018.
In Sec. \ref{sec:experiment}, we present an analysis of experimental data collected during the initial commissioning of the new device.
We use these data to evaluate the performance of the beamline and validate the models used for its design, ultimately demonstrating that experimentally relevant numbers of positrons and antiprotons can already be delivered to both the \alphatwo{} and \alphag{} experiments (see \mbox{Fig. \ref{fig:alpha-schematic}}).

\section{PARTICLE SOURCES} \label{sec:sources}
Penning traps are used extensively in experiments involving trapped ions, positrons and antiprotons \cite{Fajans-2020, Blaum-2010}.
In the ALPHA geometry, charged particles are held inside a stack of hollow cylindrical electrodes, immersed in a uniform magnetic field that is oriented along the trap axis.
Particles are confined radially due to their periodic motions in the external magnetic field (0.1 -- 3.0 T).
Longitudinal confinement is achieved by applying voltages to the electrodes to produce an electrostatic potential well along the trap axis.
\par
\begin{figure}
\centering{
\hspace{-5pt}
\includegraphics[scale=0.33, trim=17 17 20 20, clip]{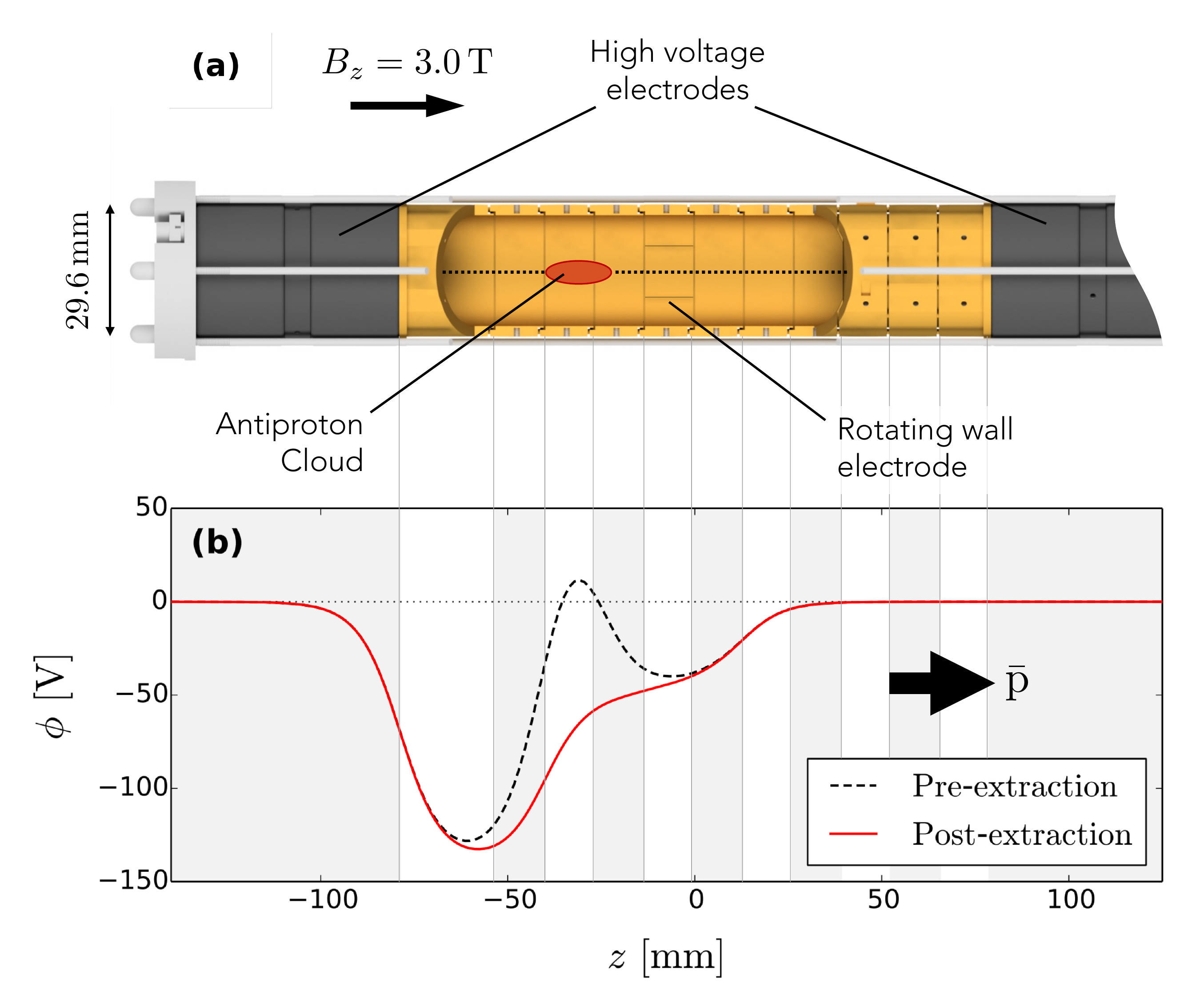}
}
\caption{Diagram showing (a) a cross section of the catching trap electrode stack and (b) the on-axis electric potential $\phi$ used to eject a \mbox{$\sim$ 50 eV} antiproton bunch. The dashed line shows the asymmetric $\sim$ 52 V potential well used to hold the \pbar{} cloud prior to extraction, and the solid line shows the potential immediately after the extraction of the \pbar{} bunch.}
\label{fig:catching-trap}
\end{figure}

Antiprotons are initially captured from the AD in a dedicated Penning trap known as the Catching Trap (CT), located on the left in \mbox{Fig. \ref{fig:alpha-schematic}}.
Captured \pbar{} bunches are sympathetically cooled by allowing them to equilibrate with an electron plasma in a strong (3.0 T) magnetic field \cite{Gabrielse-1989}.
\mbox{Figure \ref{fig:catching-trap}} shows the CT electrode stack and the asymmetric \mbox{$\sim$ 52 V} electric potential well used to confine clouds of trapped antiprotons.
A typical experimental sequence will produce a cloud of \mbox{$10^{5}$ \pbar{}}, with a radius of \mbox{0.4 mm} and a temperature of \mbox{$\sim$ 350 K} \cite{ALPHA-2017-2}.
\par
Positron bunches are accumulated using a Surko-type buffer gas trap \cite{Surko-1989, Surko-1992}, shown to the right in \mbox{Fig. \ref{fig:alpha-schematic}}.
Positrons are derived from the $\beta^{+}$ decay of a \mbox{$^{22}$Na} source with a cryogenic solid neon moderator, and formed into a dense plasma by collecting a fraction of the resulting low-energy beam in a Penning trap using an \mbox{N$_{2}$} buffer gas \cite{ALPHA-2014-1}.
The Positron Accumulator (PA) is enclosed by a long solenoid that produces a uniform magnetic field of \mbox{0.15 T}.
Typically, between $10^{6}$ and $10^{8}$ positrons are formed into a long \mbox{(150 mm)} plasma of radius \mbox{$\sim$ 1 mm} every 60 -- 90 seconds.
\par
\begin{table}
\renewcommand{\arraystretch}{1.25} 
\renewcommand{\belowcaptionskip}{-5pt} 
\centering{
\begin{tabular}{lcccc}
\hline \hline
																						&& 	Antiprotons (\pbar{}) 				&& 	Positrons (\positron{}) 	\\ \hline
Source trap																		&&	Catching trap							&&	Accumulator						\\
Magnetic field $B_{0}$ [T]												&&	3.0											&&	0.15									\\
Number of particles	$N$													&&	$10^{5}$									&&	$10^{8}$							\\
Beam radius $\sigma_{0}$ [mm]										&&	0.4											&&	1.0									\\
Source temperature $T$ [K]												&&	350											&& 	1000									\\
Larmor radius $r_{L}$ [$\mu$m]										&&	8.4											&&	6.6									\\
Beam energy $E_{\parallel}$ [eV]										&&	50											&&	50									\\
Emittance $\varepsilon$ [mm mrad]									&&	12.0											&&	51.9									\\
Magnetisation $\mathcal{L}$ [mm mrad]							&&	233											&&	3200								\\
Perveance $K$																	&&	$1.7 \times 10^{-5}$					&&	$1.0 \times 10^{-3}$			\\
\hline \hline
\end{tabular}
}
\caption{Example characteristics of antiproton and positron bunches extracted from the catching trap and positron accumulator, respectively.  The true bunch characteristics will vary depending on the experimental protocols used to prepare particles for extraction into the beamline.}
\label{table:particle-sources}
\end{table}
Particles are transferred between Penning traps in short pulses, produced by rapidly (\mbox{$\sim$ 2 $\mu$s} for antiprotons, \mbox{$\sim$ 200 ns} for positrons) modifying the electric potential along the trap axis so particles can escape in one direction, as shown in \mbox{Fig. \ref{fig:catching-trap}(b)}.
The mean energy of the beam is set by the electric potential at the centre of the trap immediately before particles begin to escape \cite{Surko-2004}.
In ALPHA, voltage breakdown limits on the Penning trap electrode cabling constrains the particle kinetic energies to \mbox{$\lesssim$ 100 eV}, therefore limiting the types of beamline that can be used to directly guide bunches between different parts of the apparatus.
\mbox{Table \ref{table:particle-sources}} summarises the properties of \pbar{} and \positron{} bunches extracted from the CT and PA.
\par
At the end of a charged particle transfer, the extracted positrons or antiprotons are captured inside the Penning trap of either the \alphatwo{} or \alphag{} experiment by rapidly \mbox{($\sim$ 0.1 $\mu$s)} applying a voltage to one of the trap electrodes to enclose the beam in an electrostatic potential well.
The maximum number of particles that can be captured from each bunch depends on the bunch length and the shape of the catching potential.
\par
Since the \pbar{} and \positron{} bunches that are transported around ALPHA have exceptionally low charge densities, their dynamics are dominated by single particle motions rather than collective (space charge) effects while in transit.
To test whether space charge forces can be neglected, we can consider how the motion of a single particle is affected by the electric potential of the surrounding bunch.
In an axial magnetic field, the radial component of the electric field within each bunch will cause particles to undergo a slow E $\times$ B rotation (at \mbox{24 kHz} for \pbar{} and \mbox{} for \positron{}) about the beamline axis, rather than causing defocusing of the beam envelope.
This rotation is generally much slower than the transit time for positrons \mbox{($\sim$ 2.8 $\mu$s)} or antiprotons \mbox{($\sim$ 120 $\mu$s)} between parts of the apparatus.
For this reason, this work will largely neglect collective effects when considering the transverse dynamics of particle bunches.
\par
Parallel to the beamline axis, space charge forces will cause bunches to elongate in the time domain.
For a \pbar{} bunch extracted with the parameters listed in \mbox{Table \ref{table:particle-sources}}, the maximum on-axis space charge potential is \mbox{$\lesssim$ 0.6 V}.
The resulting electric field will increase the spread of energies within each \pbar{} bunch by \mbox{$\sim$ 33 meV} over a typical transit time.
This increase is much smaller than the initial energy spread of each bunch (typically a few electron volts), and so longitudinal space charge forces can safely be neglected.
However, the space charge potential of a typical positron bunch is much larger (\mbox{$\sim$ 3.0 V}), causing the beam energy spectrum to widen by around \mbox{1.1 eV} while in transit.
This is comparable to the initial energy spread of each \positron{} bunch upon extraction from the accumulator, resulting in longitudinal dynamics that are dominated by collective effects which are not modelled in detail here.

\section{THEORY} \label{sec:theory}
\subsection{Semi-Analytical Methods} \label{sec:semi-analytical}
In the ALPHA Penning traps, the Larmor radius of each particle $r_{L}$ is generally much smaller than the equilibrium transverse size of the trapped plasma $\sigma_{0}$ (see \mbox{Table \ref{table:particle-sources}}), and so \pbar{} and \positron{} bunches extracted into the beamline are considered to be magnetised \cite{Piot-2014}.
While travelling at low energies through a slowly varying magnetic field, the individual particles in a magnetised beam will follow the field lines with simple motions that can be modelled accurately using the GCA.
\par
The validity of the GCA in this system can be tested using the adiabaticity parameter \cite{Hurst-2015}, which compares the typical length scale of variations in the magnetic field to that of a particle's own cyclotron motion.
In a magnetic field $\vec{B} = B \left( s \right) \hat{s}$, this is defined as
\begin{equation} \label{eq:adiabaticity}
\gamma = \frac{\tau_{c} v_{\parallel}}{B \left( s \right)} \left| \frac{\partial B}{\partial s} \right| \,,
\end{equation}
where $\tau_{c} = 2 \pi m / q B$ is a particle's instantaneous cyclotron period, $v_{\parallel}$ is its velocity along the direction of the local magnetic field $\hat{s}$, and $q$ and $m$ are its charge and mass respectively.
The co-ordinate $s$ denotes the total displacement of the beam along the magnetic field lines.
\par
In the limit where $\gamma \ll 1$, particles will robustly follow the magnetic field lines with trajectories that are well modelled by the GCA.
If $\gamma$ becomes large, particles can adopt more complex motions that are difficult to model analytically, often requiring the use of numerical calculations.
By intentionally designing the beamline so that $\gamma$ is small across all regions, we can greatly reduce the computational effort required to evaluate and optimize initial design choices.
\par
In a slowly-varying magnetic field where \mbox{$\gamma \ll 1$}, the beam envelope is path-independent and can be approximated as
\begin{equation} \label{eq:envelope-scaling}
\sigma_\mathrm{adiabatic} \left( s \right) \simeq \sigma_{0} \sqrt{\frac{B_{0}}{B \left( s \right)}} \,,
\end{equation}
where $\sigma_{0}$ and $B_{0}$ are the beam radius and magnetic field at the source, respectively.
In the regime where \mbox{$\gamma \gtrsim 1$} Eq. \ref{eq:envelope-scaling} is no longer valid, but the beam radius can still be approximated using an appropriate beam envelope equation.
\par
Ignoring acceleration effects, the envelope equation for a magnetised beam in an axial magnetic field is \cite{Reiser-1995, Piot-2014}
\begin{equation} \label{eq:beam-envelope}
\frac{\partial^{2} \sigma_\mathrm{env}}{\partial s^{2}} + k_{l}^{2} \sigma_\mathrm{env} - \frac{K}{4 \sigma_\mathrm{env}} - \frac{\varepsilon^{2} + \mathcal{L}^{2}}{\sigma_\mathrm{env}^{3}} = 0 \,,
\end{equation} 
where \mbox{$k_{l} = qB / 2 m v_{\parallel}$} is the Larmor wavenumber and $\varepsilon$ is the geometric emittance of the beam.
The perveance \mbox{$K = 2I / I_{0}$} is defined in terms of the peak beam current $I$ and Alfv\'en current $I_{0}$, and represents the defocusing effect of space charge.
In a strong magnetic field, the magnetisation $\mathcal{L}$ can be approximated as \mbox{$\mathcal{L} \, \simeq \, q B_{0} \sigma_{0}^{2} / 2 m v_{\parallel}$} \cite{Piot-2014}.
The solution to \mbox{Eq. \ref{eq:beam-envelope}} will provide a more accurate description of the beam envelope than the simple approximation given by \mbox{Eq. \ref{eq:envelope-scaling}}.
\par
In the ALPHA beamline, steering to different experiments is achieved using a region of curved magnetic field lines directly below the \alphag{} atom trap, referred to as the interconnect (see \mbox{Fig.  \ref{fig:alpha-schematic}}).
In this region, we consider a parameter $\gamma_{r}$ that is analogous to \mbox{Eq. \ref{eq:adiabaticity}} for a magnetic field that changes direction with a fixed radius of curvature, defined as
\begin{equation} \label{eq:curved-adiabatic}
\gamma_{r} = \frac{4mv_{\parallel}}{qBR} \,,
\end{equation}
where $B$ is the average magnetic field strength along the nominal beam path and $R$ is the radius of curvature of the magnetic field.
In the regime where $\gamma_{r} \ll 1$, particles will complete many cyclotron orbits as they follow the curved magnetic field lines, however if $\gamma_{r}$ becomes large particles can have more complex orbits and may not strictly follow the field lines or respect the GCA.
\par
When \mbox{$\gamma \ll 1$}, the GCA predict that charged particles following curved magnetic field lines will have curvature drifts, which displace their trajectories at right angles to both the magnetic field and its radius of curvature.
For \pbar{} and \positron{} bunches that are steered through a sharp right-angled turn while travelling towards the \alphag{} experiment, the total displacement due to this drift is approximately
\begin{equation} \label{eq:curvature-drift}
\delta x = \frac{\pi}{qB} \sqrt{\frac{E_{\parallel} m }{2}} \,,
\end{equation}
where $E_{\parallel}$ is the beam energy along the direction of the magnetic field lines.
\par
During cyclotron motion, each particle also has a magnetic moment \mbox{$\mu = E_{\perp} / B$} that is conserved adiabatically, where $E_{\perp}$ is the kinetic energy of the particle perpendicular to the magnetic field lines \cite{Chen-1984}.
In regions where \mbox{$\gamma \gtrsim 1$}, the conservation of $\mu$ can be broken, allowing particles to transfer energy between their transverse and longitudinal degrees of freedom.
In extreme cases, this can result in magnetic mirroring and particle losses when transiting from weak to strong magnetic fields.
Particle losses due to magnetic mirroring have previously been identified when transferring \positron{} bunches to the \alphatwo{} experiment.

\subsection{Numerical Simulations} \label{sec:numerical-methods}
In regions of the beamline where \mbox{$\gamma \gtrsim 1$}, the beam dynamics are poorly described by analytical models such as the GCA, and numerical calculations are required.
A range of numerical particle tracing simulations were therefore developed, and used extensively to model the trajectories of charged particles through the new beamline.
Their results were used to validate the semi-analytical models described in the previous section, and to optimize parts of the beamline where such models are not expected to be valid.
In general, numerical simulations were not used to study the dynamics of \positron{} bunches, since here $\gamma$ is consistently very small for positrons along the nominal beam paths.
\par
In a typical simulation, $10^{4}$ antiprotons were propagated from the CT up to the centre of the \alphag{} experiment.
The initial positions and velocities of beam particles were sampled from a set of distributions reflecting the source parameters in \mbox{Table \ref{table:particle-sources}}.
The displacement of each particle's guiding centre from the beamline axis was sampled from a Gaussian distribution of width \mbox{$\sigma_{0}$ = 0.4 mm}.
Particles were assigned transverse velocities from a Maxwell-Boltzmann distribution with temperature \mbox{$T$ = 350 K}, representing the thermal spread of velocities in a trapped non-neutral plasma at equilibrium.
Finally, each particle was displaced from its guiding centre along a vector corresponding to its Larmor radius in the magnetic field of the particle source.
We neglect the longitudinal structure of each \pbar{} bunch, such that all particles are initialised at a single $s$ co-ordinate along the beamline with a fixed longitudinal energy. 
\par
\begin{figure}
\renewcommand{\abovecaptionskip}{0pt}
\renewcommand{\belowcaptionskip}{-10pt}
\centering{
\includegraphics[scale=0.44, trim=15 18 5 0, clip]{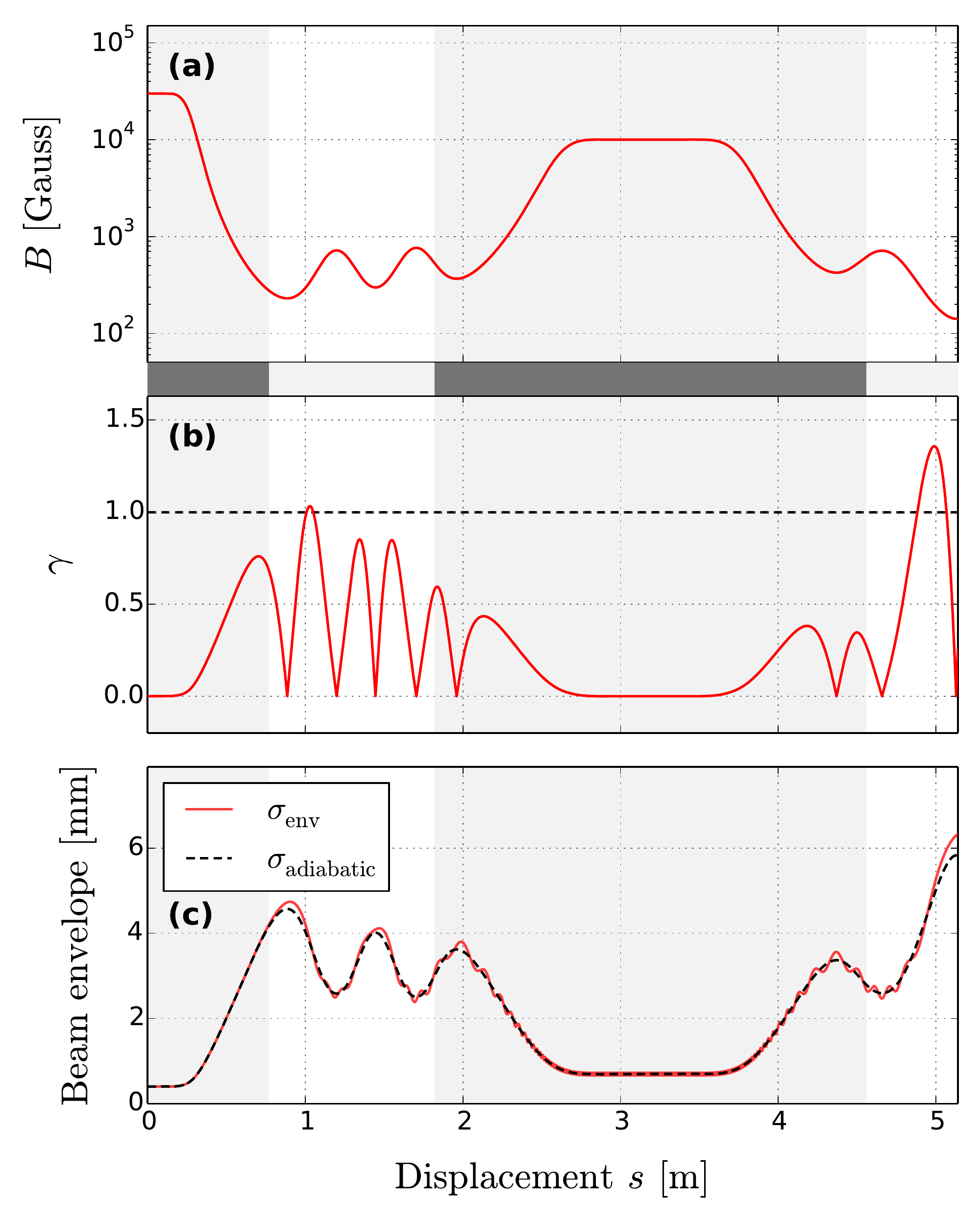}
}
\caption{Beam dynamics plots for a \mbox{50 eV} \pbar{} bunch extracted along beam path A, as shown in \mbox{Fig. \ref{fig:alpha-schematic}}. (a) Shows the on-axis magnetic field strength, while (b) shows the adiabaticity parameter $\gamma$ and (c) shows two semi-analytical solutions for the beam envelope. The horizontal axis has been shaded to match \mbox{Fig. \ref{fig:alpha-schematic}}.}
\label{fig:region-one-dynamics}
\end{figure}
As noted earlier, the magnetic field of the ALPHA experiment is formed from a complex patchwork of overlapping fields.
For numerical simulations, a magnetic field map was generated for each beamline magnet using the Biot-Savart solver of the \textsc{Opera3D} postprocessor \cite{Opera-3D}.
Each map was calculated up to the \mbox{$10^{-2}$ Gauss} contour in space (less than \mbox{$\sim$ 0.1\%} of the maximum field at the magnet's nominal operating current) to accurately model the stray fields between sections of the apparatus.
Field maps were exported on a regular \mbox{$\left( r ,\, z \right)$} grid with a resolution of 2.5 mm, and the field between grid points was evaluated using cubic interpolation.
Each two-dimensional field map was located and aligned within the cartesian co-ordinate system of the simulation using an appropriate set of co-ordinate transformations.
At any point along the beamline, the total magnetic field can be found by scaling and superimposing these maps according to the current in each magnet.
\par
We used the leapfrog (Boris) algorithm \cite{Qin-2013} to solve the equation of motion for each particle.
In a pure magnetostatic field this algorithm conserves energy exactly, making it ideal for following particles with fast periodic motions (e.g. the cyclotron motion) over long timescales.
In the ALPHA beamline, antiprotons can have cyclotron frequencies of up to \mbox{45 MHz}, while a typical particle transfer will last \mbox{$\sim$ 120 $\mu$}s.
A simulation timestep of \mbox{$10^{-9}$ seconds} was found to give a good compromise between maximising convergence and limiting the computational requirements.

\section{DESIGN} \label{sec:design}
The following section describes the elements that make up the ALPHA beamline.
To differentiate between the new instrument and pre-existing parts of the ALPHA apparatus, \mbox{Fig. \ref{fig:alpha-schematic}} has been divided into two regions.
\par
In \mbox{Sec. \ref{sec:region-one}} we discuss region one, which includes both the CT and the \alphatwo{} experiment, while in \mbox{Sec. \ref{sec:region-two}} we consider region two, which includes the new beamline described here.
The existing beamline (region one) has been used extensively since 2012 to transfer antiprotons to \alphatwo{}, and is unmodified by this upgrade.
In \mbox{Sec. \ref{sec:beam-capture}} we consider the longitudinal dynamics of \pbar{} and \positron{} bunches, and outline mechanisms that may cause particle losses inside the \alphag{} Penning trap.
\begin{figure}
\renewcommand{\abovecaptionskip}{0pt}
\renewcommand{\belowcaptionskip}{-5pt}
\centering{
\includegraphics[scale=0.32, trim=15 0 0 0, clip]{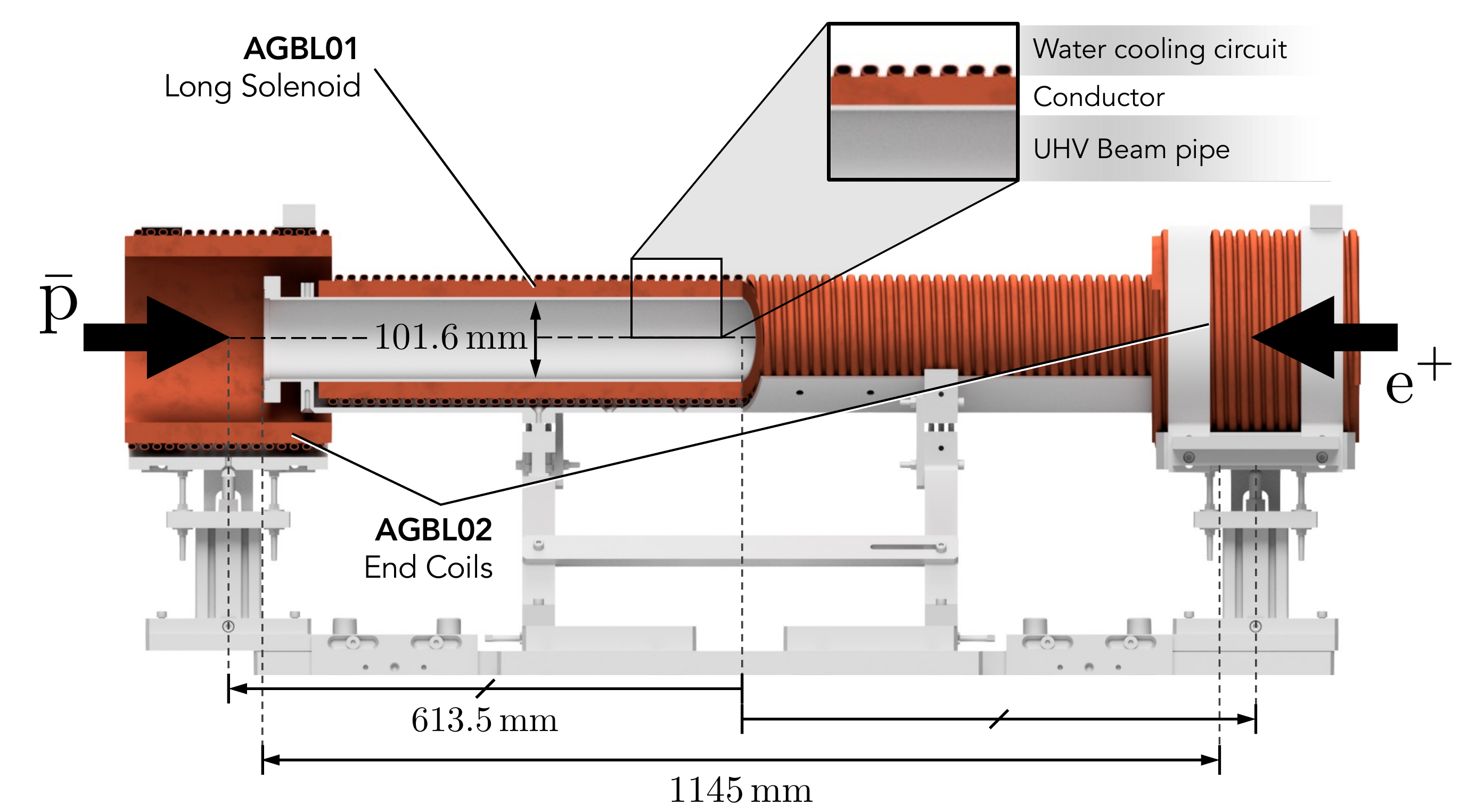}
}
\caption{Diagram showing the cross-section of an ALPHA beamline module. Copper conductors and water cooling circuits are highlighted in orange, while the UHV beam pipe and its surrounding support structures are shown in grey.}
\label{fig:beamline-module}
\end{figure}

\subsection{Region One} \label{sec:region-one}
As shown in \mbox{Fig. \ref{fig:alpha-schematic}}, \pbar{} bunches transferred to \alphag{} from the CT must initially travel through the pre-existing \alphatwo{} experiment.
\mbox{Figure \ref{fig:region-one-dynamics}(a)} shows the magnetic field strength along this section of the beamline, and \mbox{Fig. \ref{fig:region-one-dynamics}(b)} shows the value of $\gamma$ for a \mbox{50 eV} \pbar{} beam propagating through this field from the CT.
The horizontal axis of \mbox{Fig. \ref{fig:region-one-dynamics}} is labelled as beam path A in Fig. \ref{fig:alpha-schematic}.
In this region, the magnetic field is sufficiently strong and slowly varying that \mbox{$\gamma \lesssim 1$} along the full length of the beam path shown in Fig. \ref{fig:alpha-schematic}.
\par
\mbox{Figure \ref{fig:region-one-dynamics}(c)} shows two models for the \pbar{} beam envelope, obtained by evaluating \mbox{Eq. \ref{eq:envelope-scaling}} \mbox{($\sigma_\mathrm{adiabatic}$)} and solving \mbox{Equation \ref{eq:beam-envelope}} \mbox{($\sigma_\mathrm{env}$)} for a \pbar{} beam extracted from the CT.
The beam envelope equation was solved using a fourth-order Runge-Kutta integrator, for a monoenergetic beam with the initial conditions listed in \mbox{Table \ref{table:particle-sources}}.
Since the magnetic field is strong and relatively uniform within this region, the solutions to both equations are expected to be good approximations to the real transverse size of the beam.
\begin{figure}[h]
\renewcommand{\abovecaptionskip}{-5pt}
\renewcommand{\belowcaptionskip}{-5pt}
\centering{
\includegraphics[scale=0.44, trim=15 18 5 0, clip]{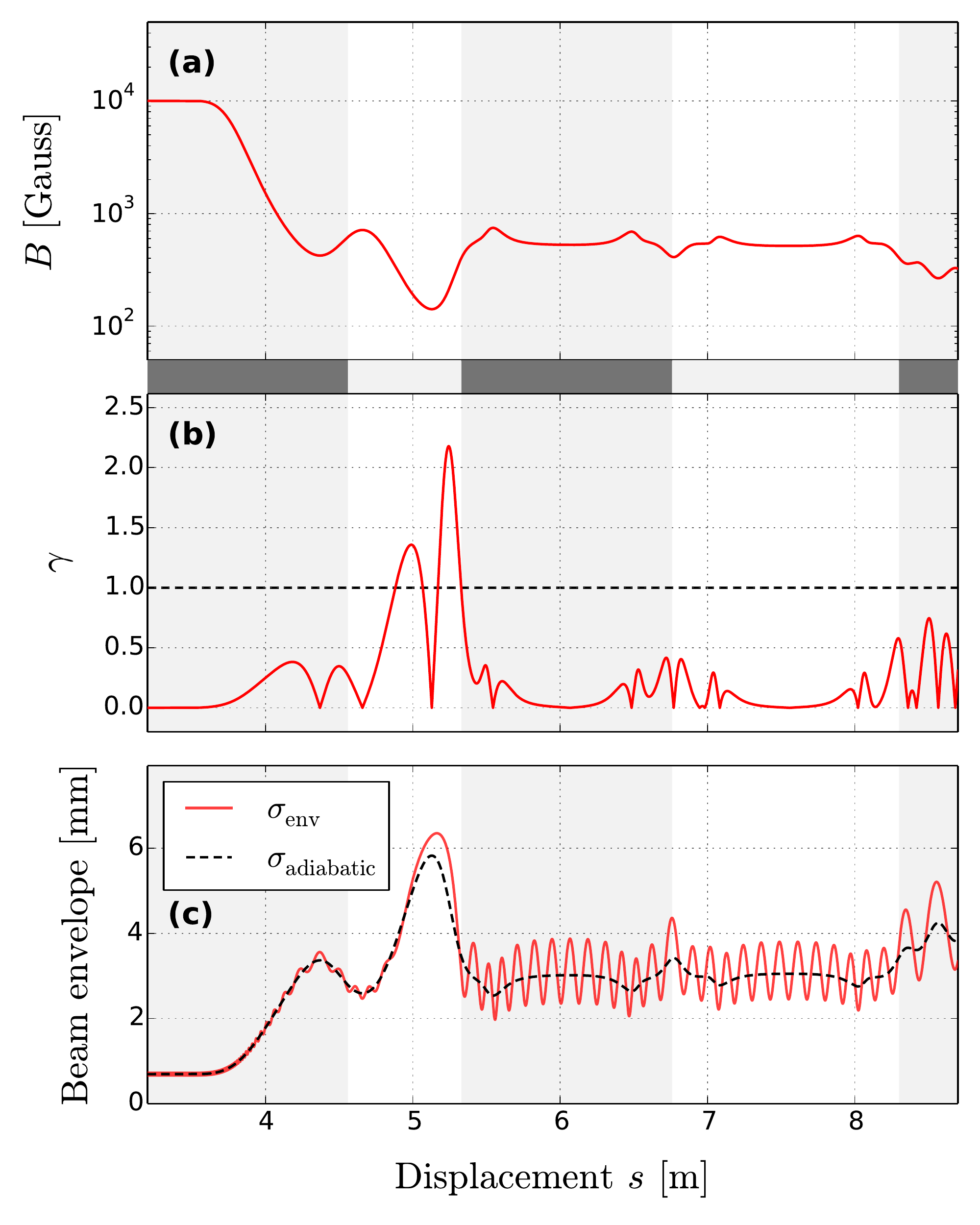}
}
\caption{Beam dynamics plots for a \mbox{50 eV} \pbar{} bunch extracted along beam path B, as shown in \mbox{Fig. \ref{fig:alpha-schematic}}. (a) Shows the on-axis magnetic field strength, while (b) shows the adiabaticity parameter $\gamma$ and (c) shows two semi-analytical solutions for the beam envelope. The horizontal axis has been shaded to match \mbox{Fig. \ref{fig:alpha-schematic}}.}
\label{fig:region-two-dynamics}
\end{figure}

\subsection{Region Two} \label{sec:region-two}
The new ALPHA beamline (labelled as region two in \mbox{Fig. \ref{fig:alpha-schematic}}) can be divided into six independent sections, with each section containing one of three types of elements.
At present, the beamline is comprised of three straight beamline modules, two diagnostics stations and the interconnect.
In the following section, we outline the design of each beamline element and discuss the transverse dynamics of \pbar{} bunches propagating along beam path B in \mbox{Fig. \ref{fig:alpha-schematic}}.
\begin{table*}
\setlength{\tabcolsep}{10pt}
\renewcommand{\arraystretch}{1.1} 
\renewcommand{\abovecaptionskip}{10pt}
\renewcommand{\belowcaptionskip}{-5pt}
\centering{
\begin{tabular}{lccccccc}
\hline \hline
\textbf{Identifier} &	 \multicolumn{2}{c}{\textbf{Radius [mm]}} & \textbf{Length} 	&	 \multicolumn{3}{c}{\textbf{Nominal Current [A]}}				& \textbf{Peak Field} 	\\ 
\hspace{50pt}	 			&		Inner			&		Outer 			& 	\textbf{[mm]}					& Antiprotons 			& Positrons 			& Positrons 						&	\textbf{[Gauss]}		\\
									&						&		 					& 											& (\alphag{}) 			& (\alphag{}) 		& (\alphatwo{}) 	&	\\	 \hline					
\multicolumn{8}{l}{\textbf{Beamline Modules}} \\
AGBL01							&		50.8			& 		70.8				&		1012.6										&		$+11.0$			&	$+8.0$ 			& 		$+8.0$				&	730	\\	
AGBL02						&		100.0		& 		125.0			&		250.0										&		$+11.0$			&	$+8.0$  			& 		$+8.0$				&	680 \\	
\multicolumn{8}{l}{\textbf{Diagnostics Stations}} \\
AGBL03						&		132.5		& 		171.9				&		41.1											&		$+15.0$			&	$+15.0$ 			&		$+5.0$				&	274	\\	
\multicolumn{8}{l}{\textbf{Interconnect}} \\
AGBL04-US					&		77.5			& 		105.0			&		95.0											&		$+15.1$			& $+10.6$ 			&		$+16.0$			&	640	\\
AGBL04-DS					&		77.5			& 		105.0			&		95.0											&		$-15.1$				& $-10.6$ 			&		$+16.0$			&	640	\\
AGBL05						&		90.0			& 		120.0			&		100.0										&		$+15.8$			& $+9.8$ 				& 		$0.0$				&	680	\\
AGBL06						&		130.0		& 		160.0			&		60.0											&		$+13.2$			& $0.0$ 				& 		$0.0$				&	270	\\
AGBL07 (Inner)				&		117.5			& 		142.5			&		60.0											&		$+16.6$			& $-15.0$ 			& 		$0.0$				&	310	\\
AGBL07 (Outer)			&		145.0		& 		170.0			&		60.0											&		$+16.6$			& $-15.0$ 			& 		$0.0$				&	260	\\
\multicolumn{8}{l}{\textbf{Transfer Coils}} \\
AGBL08						&		302.5		&		350.0			&		100.0										&		$+9.5$				&	$+9.5$				& 		$0.0$				&	226	\\
\hline \hline
\end{tabular}
}
\caption{Design specifications for the ALPHA beamline magnets. Nominal operating currents are given for each of the three main beamline configurations, used to transfer \pbar{} bunches to the \alphag{} experiment or \positron{} bunches to either the \alphatwo{} or \alphag{} experiment. Peak magnetic fields were calculated for the \pbar{} beamline configuration. The direction of current in each magnet is defined relative to its polarity in the \pbar{} beamline configuration, which ensures continuity of the magnetic field between the CT and \alphag{}.}
\label{table:magnets}
\end{table*}
\subsubsection{Straight Beamline Modules}
\mbox{Figure \ref{fig:beamline-module}} shows the cross section of a single beamline module.
All three beamline modules share the same basic geometry, and act as long guiding channels for \pbar{} and \positron{} bunches.
Each module consists of a central long solenoid (AGBL01), enclosed at either end by a pair of shorter solenoids (AGBL02) that are powered in series and can be positioned along the axis of the beamline.
The polarities of the magnets are chosen to ensure continuity of the magnetic field lines along the \pbar{} and \positron{} beam paths.
The specifications for each type of magnet are summarised in \mbox{Table \ref{table:magnets}}.
The identifiers in \mbox{Table \ref{table:magnets}} refer to magnet designs that may be used in multiple places along the beamline.
\par
For the beamline to be modular, lengths of beam pipe are separated using gate valves and bellows so that each section can be isolated and positioned independently.
These components create drift spaces between the beamline magnets where $\gamma$ can approach 1 for antiprotons, resulting in emittance growth and expansion of the \pbar{} beam envelope.
To prevent this, the two end coils (AGBL02) are moved outwards from the centre of each beamline module (as shown in \mbox{Fig. \ref{fig:beamline-module}}), creating a significant stray field and maximising the magnetic field between sections of the beamline.
All data and simulations presented in this work assume a beamline configuration where the end coils are fully deployed, as shown in \mbox{Figs.  \ref{fig:alpha-schematic} and \ref{fig:beamline-module}}.
\par
\mbox{Figure \ref{fig:region-two-dynamics}(a)} shows the magnetic field strength along beam path B in \mbox{Fig. \ref{fig:alpha-schematic}}, from the centre of the \alphatwo{} experiment up to the interconnect.
As shown in \mbox{Fig. \ref{fig:region-two-dynamics}(b)}, \mbox{$\gamma \lesssim 1$} for antiprotons along the majority of the \pbar{} beam path.
However, several regions still exist where the magnetic field is very weak or inhomogeneous, due to mechanical constraints imposed by pre-existing parts of the ALPHA apparatus.
\par
\mbox{Figure \ref{fig:region-two-dynamics}(c)} shows two semi-analytical models for the \pbar{} beam envelope, obtained using the same methods as \mbox{Fig. \ref{fig:region-one-dynamics}(c)}.
Both models are broadly in agreement along the length of the new beamline.
However, after passing through a region where \mbox{$\gamma \gtrsim 1$}, the solution to the beam envelope equation begins to oscillate, resulting in a slightly larger envelope throughout the apparatus.
The $\sigma_\mathrm{env}$ solution agrees closely with the results of an equivalent particle tracing simulation, shown in \mbox{Fig. \ref{fig:numerical-dynamics}(c)}.
\begin{figure}[h]
\vspace{5pt}
\centering{
\hspace{-10pt}
\includegraphics[scale=0.27]{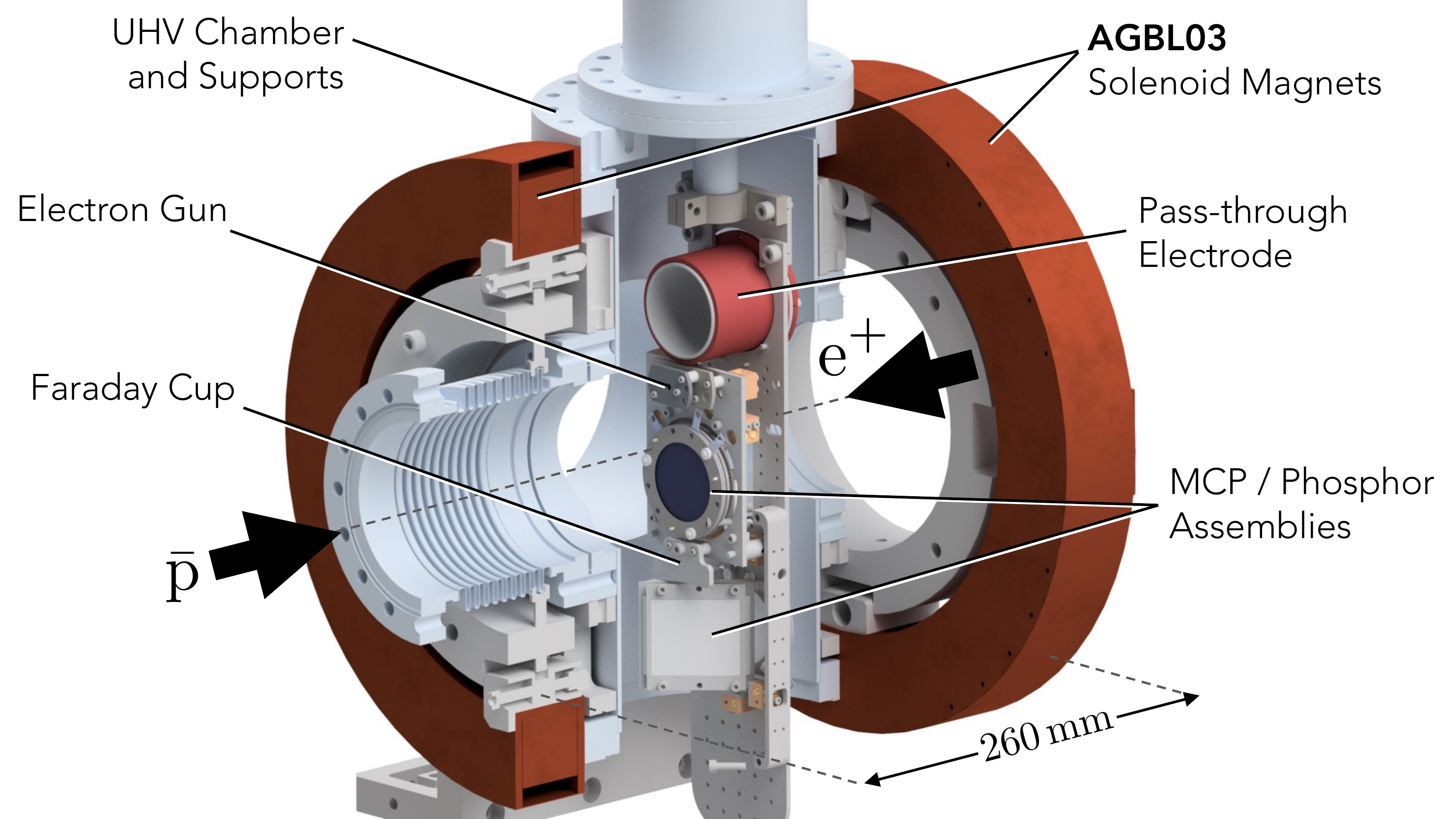}
}
\caption{Simplified schematic showing a cross section of the positron diagnostics station (PDS). The central UHV chamber and surrounding support structures are shown in light blue, while magnet windings are highlighted in orange.}
\label{fig:diagnostic-station}
\end{figure}
\subsubsection{Diagnostics Stations}
The new beamline incorporates four diagnostics stations, where measurement devices can be positioned within the ultra-high vacuum (UHV) space of the apparatus using a linear translator \cite{ALPHA-2014-1}.
These instruments are primarily used to characterise \pbar{} and \positron{} bunches in transit along the new beamline.
The locations of the beamline diagnostics station (BDS) and positron diagnostics station (PDS) are marked in \mbox{Fig. \ref{fig:alpha-schematic}}.
\par
\begin{figure*}
\renewcommand{\belowcaptionskip}{0pt}
\centering{
\hspace{-10pt}
\includegraphics[scale=0.37, trim=0 0 0 0, clip]{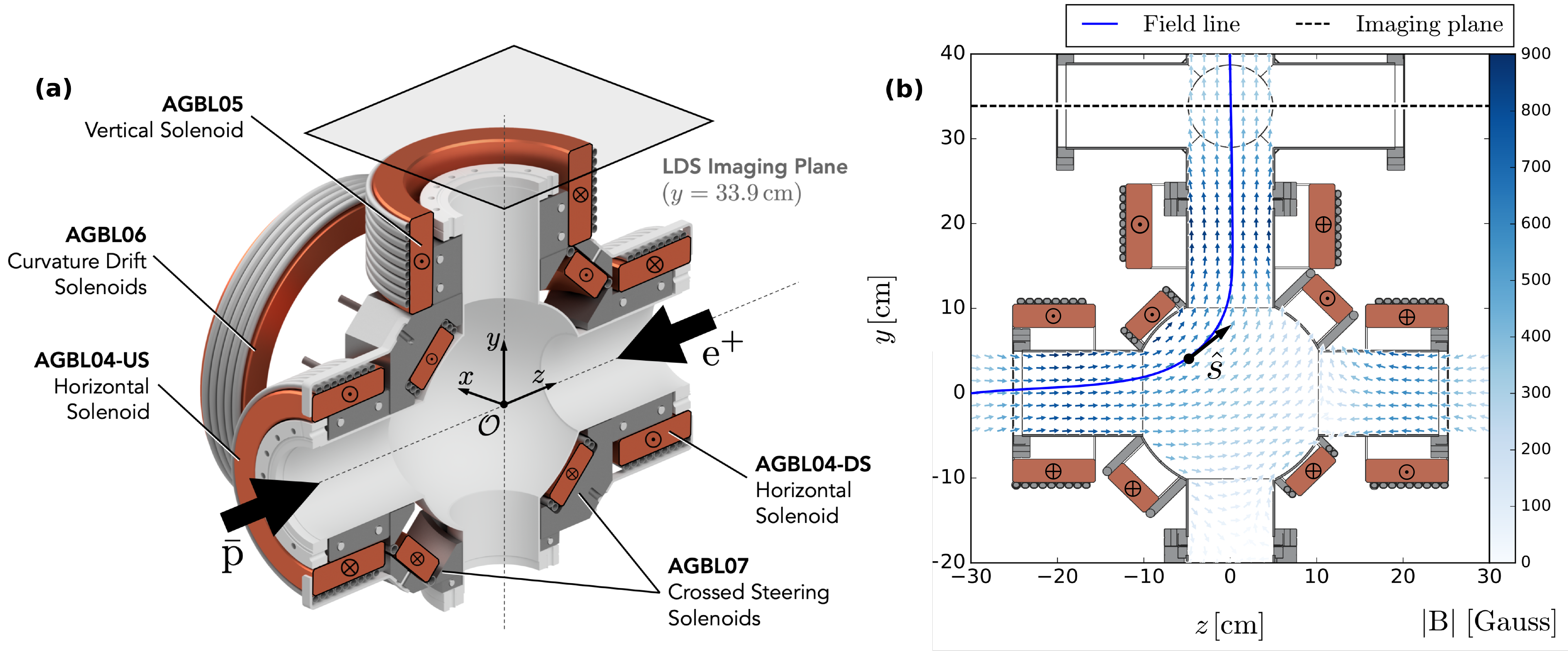}
}
\caption{(a) Simplified schematic showing a cross section of the interconnect.The magnet windings are highlighted in orange, while the UHV chamber, cooling circuits and magnet support structures are shown in grey. One of the AGBL06 magnets is not visible in this cutaway view. Crosses (dots) indicate current flowing into (out of) the shown cross-section of each magnet. (b) Quiver plot showing the strength (colour) and direction (arrow orientation) of the magnetic field within the midplane of the interconnect. The blue line shows a magnetic field line traced from the horizontal axis of the experiment, while the dashed line indicates the LDS imaging plane.}
\label{fig:interconnect}
\end{figure*}
Both stations share the same basic geometry, shown as a cross-section (specifically for the PDS) in \mbox{Fig. \ref{fig:diagnostic-station}}.
Two solenoids (AGBL03) are arranged around the midplane of each station in a Helmholtz-like configuration, at a separation of \mbox{$\pm$130 mm}.
Together, these magnets generate a uniform magnetic field of \mbox{$\sim$ 242 Gauss} around devices that are inserted into the beamline, and also focus \pbar{} and \positron{} bunches that are transferred directly through the station.
\mbox{Table \ref{table:magnets}} lists the design specifications for the two solenoids.
\par
In addition, two sets of diagnostic devices are mounted directly above and below the vertical \alphag{} experiment.
The locations of the lower diagnostics station (LDS) and upper diagnostics station (UDS) are marked in \mbox{Fig. \ref{fig:alpha-schematic}}.
Two large transfer solenoids (AGBL08) are mounted directly above the LDS to boost the total magnetic field to \mbox{$\sim$ 230 Gauss} around instruments inserted at this location.
\mbox{Table \ref{table:magnets}} lists the design specifications for these magnets.

\subsubsection{Interconnect Magnets} \label{sec:interconnect}
The interconnect (labelled in \mbox{Fig. \ref{fig:alpha-schematic}}) is used to steer charged particles to either the \alphatwo{} or \alphag{} experiment.
As noted in \mbox{Sec. \ref{sec:theory}}, beam steering within the interconnect is achieved using a region of tightly curved magnetic field lines.
To transfer both \pbar{} and \positron{} bunches into the \alphag{} experiment, the interconnect magnets must be capable of operating in multiple configurations that connect either the CT or PA to the magnetic field of the vertical atom trap.
The interconnect must also be able to operate in a mode that allows positrons to pass freely along the horizontal beamline for operation of the \alphatwo{} experiment. 
\par
\mbox{Figure \ref{fig:interconnect}(a)} shows a cross section of the interconnect, while \mbox{Table \ref{table:magnets}} lists the specifications for its seven magnet windings.
While each of these magnets is powered independently, the AGBL06 and AGBL07 magnets are generally operated as two pairs of magnets with matching currents.
To switch between the various beamline configurations (see \mbox{Table \ref{table:magnets}}), the AGBL07 magnets are connected to bipolar power circuits that allow their polarities to be changed between particle transfers.
\par
When only the two horizontal solenoids (AGBL04) are powered with the same polarity, the magnetic field of the interconnect points along the $z$ axis of \mbox{Fig. \ref{fig:interconnect}(a)}.
This configuration is used to transfer \positron{} bunches into the \alphatwo{} experiment, with the interconnect acting as a short focusing solenoid.
At present, simultaneous operation of the \alphatwo{} and \alphag{} experiments is not possible, as this would require either \pbar{} or \positron{} bunches to pass through an inversion in the magnetic field of the beamline.
Instead, the polarities of the beamline magnets are configured to ensure continuity of the magnetic field lines either between the CT and \alphag{}, or the PA and \alphatwo{}.
\par
\mbox{Figure \ref{fig:interconnect}(b)} shows the magnetic field used to steer antiprotons into the \alphag{} experiment.
\mbox{Table \ref{table:magnets}} lists the magnet currents that are used to generate this magnetic field, and the relative polarities of the different interconnect windings.
In this configuration, the two crossed solenoids (AGBL07) are powered in such a way that their total magnetic field is oriented along the $z$ axis of \mbox{Fig. \ref{fig:interconnect}(a)}, boosting the field strength around the centre of the interconnect.
\par
To test whether individual particles will follow the magnetic field lines, we can evaluate \mbox{Eq. \ref{eq:curved-adiabatic}} for a simplified version of the interconnect.
For a magnetic field with \mbox{$B = 500$ Gauss} and \mbox{$R$ = 250 mm}, we find that \mbox{$\gamma_{r} \left(\mathrm{e}^{+}\right) \simeq 7 \times 10^{-3}$} and \mbox{$\gamma_{r} \left(\mathrm{\bar{p}}\right) \simeq 0.33$}.
This implies that positrons will undergo thousands of cyclotron orbits as they sample the magnetic field of the interconnect, while antiprotons will only complete three oscillations.
\par
Assuming that charged particles robustly follow the magnetic field lines, we can also estimate their curvature drifts using \mbox{Eq. \ref{eq:curvature-drift}}.
Using the same parameters as above, we find that \positron{} bunches are displaced by around \mbox{0.8 mm} along the $x$ axis of \mbox{Fig. \ref{fig:interconnect}(a)}.
However, antiprotons will have much larger curvature drifts of up to \mbox{$\sim$ 32 mm} in the same direction.
\par
To correct the large curvature drifts of \pbar{} bunches, two additional magnets (AGBL06) are installed on either side of the interconnect.
These solenoids generate a magnetic field parallel to the $x$ axis of \mbox{Fig. \ref{fig:interconnect}(a)}.
When superimposed over the fields of the other interconnect magnets, this causes the field lines to be deflected along the $-x$ direction.
In the regime where \mbox{$\gamma_{r} \lesssim 1$}, \pbar{} and \positron{} bunches will follow the magnetic field lines in a direction that opposes their own curvature drifts.
\par
\begin{figure}[h!]
\renewcommand{\abovecaptionskip}{0pt}
\centering{
\hspace{-20pt}
\includegraphics[scale=0.37, trim=15 15 0 10, clip]{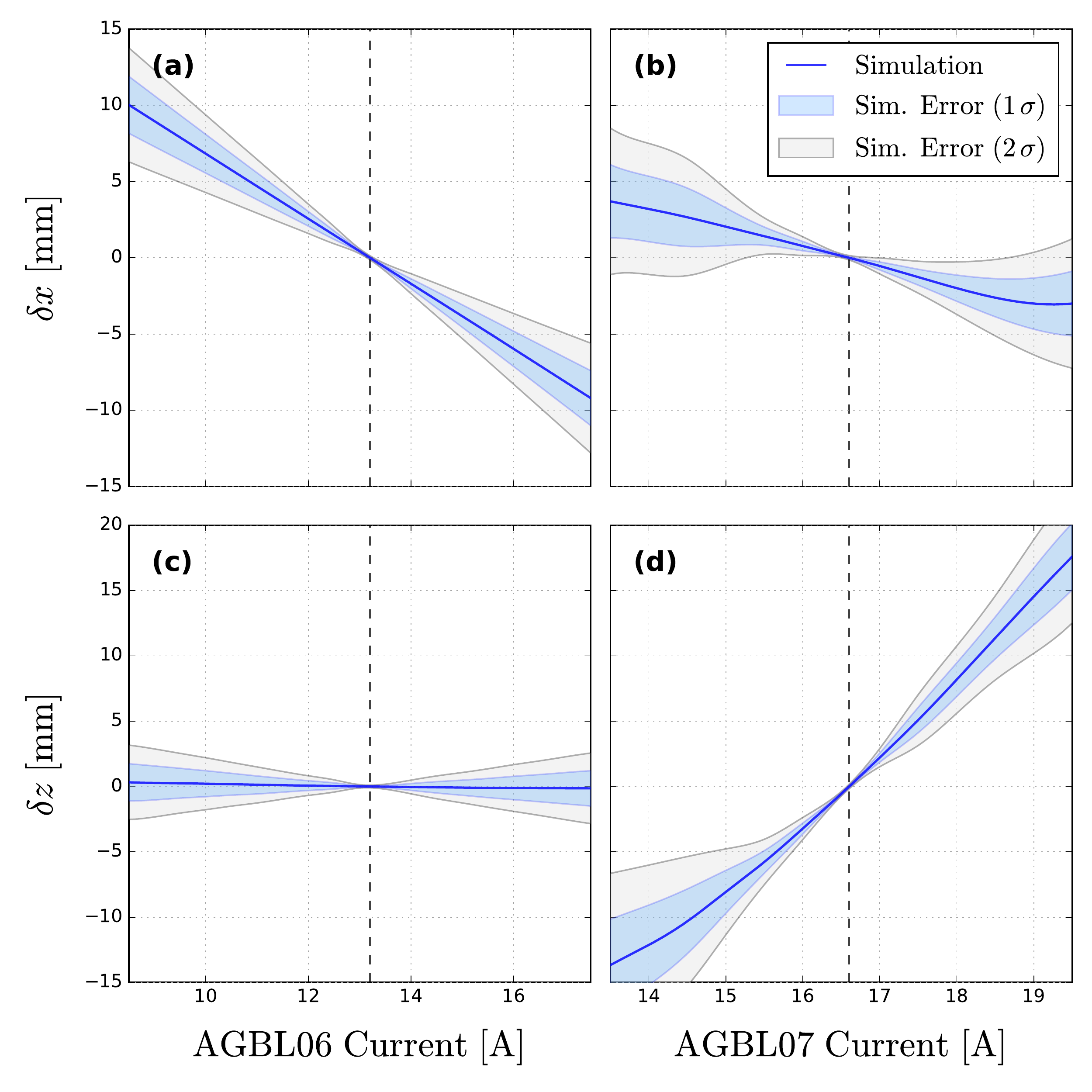}
}
\caption{Simulated steering scans showing the \pbar{} beam position at the LDS as a function of the AGBL06 and AGBL07 interconnect magnet currents. The beam position has been separated into orthogonal components along the $x$ and $z$ axes of \mbox{Fig. \ref{fig:interconnect}(a)}. In these simulations, the \alphag{} external solenoid is powered with a nominal magnetic field of \mbox{1.0 T.} The shaded regions indicate  uncertainties in the beam position due to typical mechanical errors along the full \pbar{} beamline.}
\label{fig:steering-sim}
\end{figure}
Like the \alphatwo{} experiment, \alphag{} is enclosed by a large \mbox{(510 mm bore)} superconducting solenoid that generates a magnetic field of \mbox{1.0 T}, in this case along the vertical axis of the experiment (see \mbox{Fig. \ref{fig:alpha-schematic}}).
This field is primarily needed to confine clouds of charged particles inside the \alphag{} Penning traps prior to antihydrogen synthesis.
Around the centre of the interconnect, this solenoid produces a stray field of $\sim$ 50 Gauss along the $y$ axis of \mbox{Fig. \ref{fig:interconnect}(a)}, which significantly alters the steering of charged particles through this region of the beamline.
\par
Numerical particle tracing simulations were used to model the trajectories of \pbar{} bunches through the interconnect.
\mbox{Figure \ref{fig:interconnect}(b)} shows a magnetic field line traced through the interconnect from the horizontal axis of the beamline, indicating the trajectory of an ideal \pbar{} beam with \mbox{$\gamma_{r} \lesssim 1$.}
\par
\begin{figure}
\renewcommand{\belowcaptionskip}{-10pt}
\centering{
\includegraphics[scale=0.44, trim=20 15 10 10, clip]{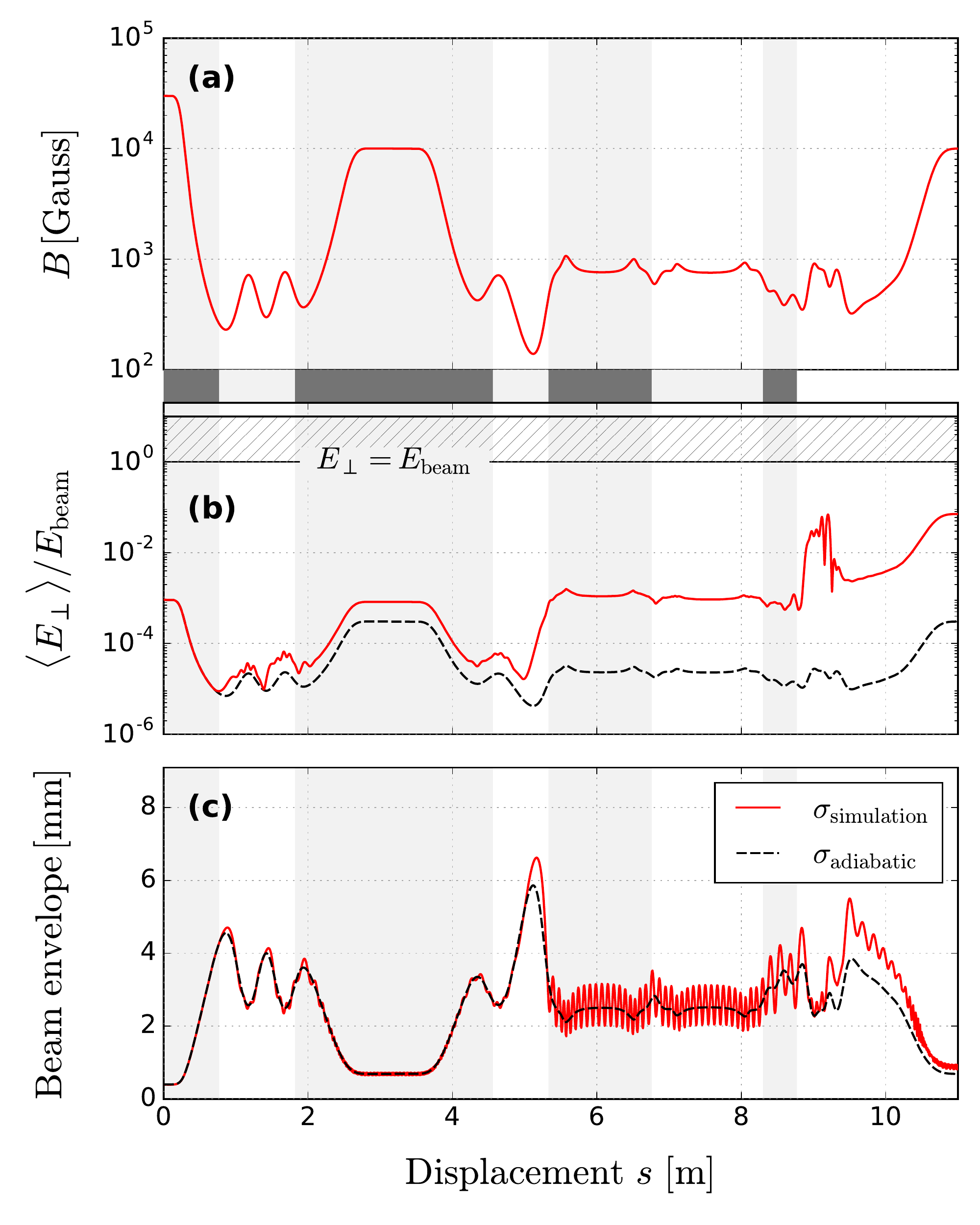}
}
\caption{Numerical simulation showing the properties of a \mbox{50 eV} \pbar{} bunch extracted from the CT to the \alphag{} experiment, as a function of its longitudinal displacement $s$. (a) Shows the magnetic field strength $B$, while (b) shows the mean \pbar{} transverse energy $\langle E_{\perp} \rangle$ as a fraction of the beam energy $E_\mathrm{beam}$. (c) Shows the simulated transverse beam envelope $\sigma_\mathrm{simulation}$ alongside \mbox{Eq. \ref{eq:envelope-scaling}}, which has been evaluated for the magnetic field shown in (a).}
\label{fig:numerical-dynamics}
\end{figure}
Adjustments to the beam trajectory can be made by tuning the currents in each of the seven magnet windings.
\mbox{Figure \ref{fig:steering-sim}} shows the simulated \pbar{} beam position at the LDS as a function of currents in select interconnect magnets.
Each data point corresponds to a single simulation using the setpoints listed in \mbox{Table \ref{table:magnets}}, while independently varying the currents in either the AGBL06 or AGBL07 magnets about their nominal values (shown as dashed lines).
The shaded bands in \mbox{Fig. \ref{fig:steering-sim}} represent the uncertainty in the simulated beam position due to typical mechanical errors along the full \pbar{} beamline.
We estimate this uncertainty as the standard deviation of 200 Monte Carlo replica simulations, where the position and orientation of each magnet have been adjusted at random within its expected mechanical tolerances.

\subsection{Longitudinal Dynamics} \label{sec:beam-capture}
As discussed in \mbox{Sec. \ref{sec:sources}}, the number of particles that are captured after a transfer depends strongly on the longitudinal spatial structure of the beam upon arrival in either the \alphatwo{} or \alphag{} Penning trap.
While in transit, the longitudinal emittances of antiproton bunches can increase due to energy transfer between their transverse and longitudinal degrees of freedom.
\par
\mbox{Figure \ref{fig:numerical-dynamics}(b)} shows the mean transverse energy $\langle E_{\perp} \rangle$ of a simulated \pbar{} beam as a fraction of the beam energy.
The horizontal axis indicates the longitudinal position of the beam, extending from the CT up to the magnetic centre of the \alphag{} experiment.
For comparison, the transverse energy ratio of a \pbar{} beam with a strictly conserved magnetic moment is shown as a dashed line.
In regions where the magnetic field is very weak or inhomogeneous (see \mbox{Fig. \ref{fig:numerical-dynamics}}(a)) $\mu$ is not conserved, and particles can transfer energy between their transverse and longitudinal degrees of freedom.
The transverse energy of the simulated \pbar{} beam therefore tends to increase while in transit along the beamline.
\par
Upon reaching the \alphag{} experiment, each antiproton has transferred an average of \mbox{$\sim$ 3 eV} of its initial longitudinal energy into its cyclotron motion.
The amount of energy that each particle moves into its cyclotron motion depends on its exact trajectory along the beamline, resulting in a distribution of transverse energies around this mean value.
If $\gamma$ had not minimised along the \pbar{} beam path, some antiprotons would enter the \mbox{1.0 T} magnetic field of the \alphag{} experiment with very large transverse energies ($E_{\perp} \sim$ \mbox{50 eV}) resulting in particle losses due to magnetic mirroring.
\par
In simulations of the antiproton beamline, every particle was launched from the same point along the beamline with a fixed longitudinal energy.
Upon arriving in the \alphag{} experiment, the beam had developed a parallel energy spread of \mbox{1.5 eV} and a bunch length of \mbox{$\sim$ 0.41 $\mu$s}.
The minimum bunch length that can be delivered to the \alphag{} Penning trap is therefore determined by mixing between the beam's transverse and longitudinal degrees of freedom while in transit along the beamline.
\par
Numerical simulations were also used to investigate the capture of \pbar{} and \positron{} bunches inside the \alphag{} experiment at the end of a particle transfer.
In these simulations, we modelled the trajectories of individual particles as they move into a time-dependent electric potential along the axis of the experiment's Penning trap.
The vacuum electric potential was found by analytically solving the Laplace equation for a hollow conducting cylinder, and superimposing this solution along the length of the trap to model each electrode.
\par
\begin{figure}
\renewcommand{\abovecaptionskip}{10pt}
\renewcommand{\belowcaptionskip}{0pt}
\centering{
\includegraphics[scale=0.42, trim=18 15 10 0, clip]{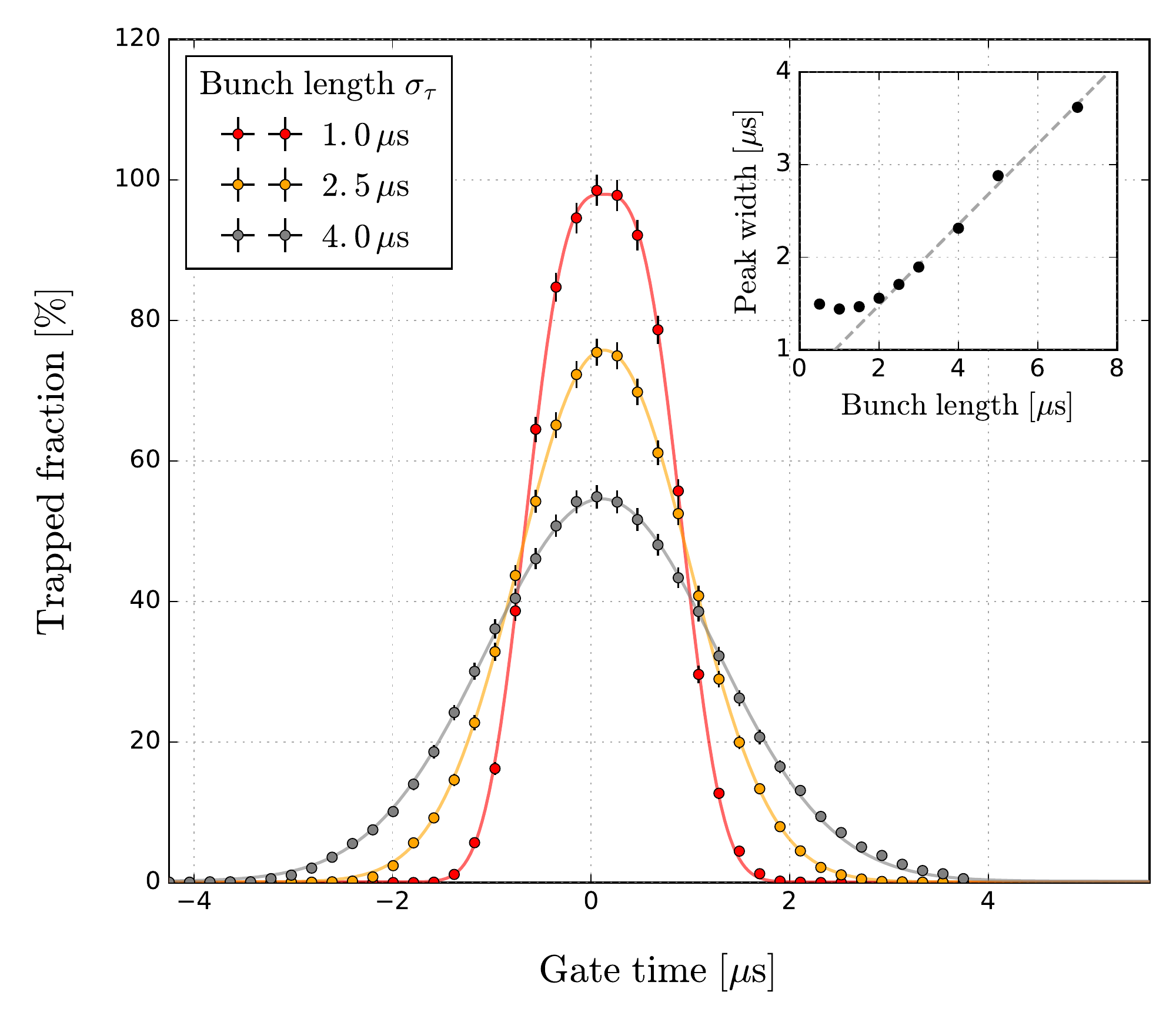}
}
\caption{Simulation showing the fraction of antiprotons captured from each \pbar{} bunch as a function of the Penning trap gate time $t_\mathrm{gate}$. Curves are shown for a range of bunch lengths between $1\,\mu\mathrm{s} \leq \sigma_{\tau} \leq 4\,\mu\mathrm{s}$. The horizontal axis has been shifted relative to the optimal gate time. The inset plot shows the widths of the fitted curves as a function of bunch length.}
\label{fig:gate-scan-sim}
\end{figure}
In each simulation, $10^{5}$ \pbar{} were initialised far outside the trap volume with a normal distribution in the time domain, and assigned parallel energies from a Gaussian distribution of width \mbox{$\sim$ 1.5 eV} centered at \mbox{50 eV}.
We define the initial bunch length $\sigma_{\tau}$ as the time interval that encloses \mbox{95 \%} of beam particles in the time domain.
After a simulated amount of time (the `gate time' $t_\mathrm{gate}$, since \pbar{} extraction from the CT), the electrode voltages were changed to capture particles inside a \mbox{105 mm} long, \mbox{140 V} electrostatic potential well.
We count the number of particles that remain inside this well after \mbox{2 ms} of simulation time to determine the number of \pbar{} captured from each bunch.
\par
\begin{figure*}
\renewcommand{\abovecaptionskip}{-3pt}
\renewcommand{\belowcaptionskip}{-7pt}
\centering{
\includegraphics[scale=0.42, trim=25 0 15 0, clip]{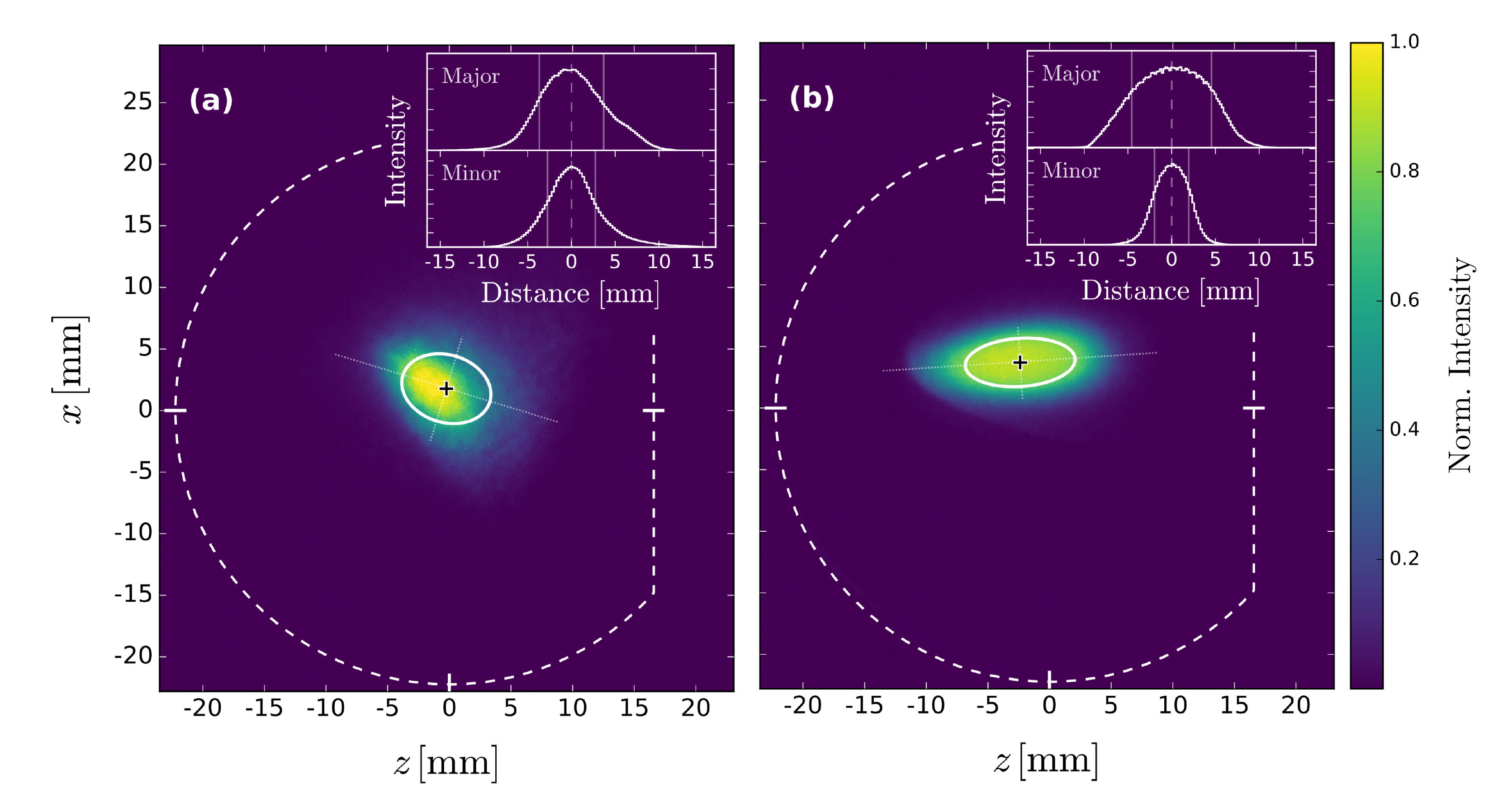}
}
\caption{MCP images showing the transverse beam profiles of (a) antiprotons and (b) positrons measured at the LDS. The dashed line shows the visible active area of the MCP. The solid white lines show the elliptical (one sigma) beam envelopes, with major and minor axes indicated by dotted lines. The inset figures show the beam intensity integrated along the major and minor axes of each ellipse. The directions of the $x$ and $z$ axes are defined in \mbox{Fig. \ref{fig:interconnect}(a)}.}
\label{fig:lds-mcp-images}
\end{figure*}
\mbox{Figure \ref{fig:gate-scan-sim}} shows how the simulated number of antiprotons captured from a \mbox{50 eV} \pbar{} bunch varies as a function of the gate time and initial bunch length.
To avoid significant particle losses, the bunch length inside the \alphag{} experiment must be limited to \mbox{$\sigma_{\tau} \lesssim$ 1 $\mu$s}.
For positrons, the maximum bunch length is on the order of \mbox{0.25 $\mu$s}.
The \pbar{} bunch length that is delivered to the \alphag{} Penning trap can be reduced by minimising the initial spread of longitudinal energies within each bunch upon extraction from the CT.
However,  the longitudinal dynamics of \positron{} bunches are dominated by space charge forces (see \mbox{Sec. \ref{sec:sources}}) that cannot easily be mitigated by tuning the initial bunch parameters.

\section{EXPERIMENTAL PERFORMANCE} \label{sec:experiment}
\subsection{Beam Diagnostics}
In this section, we evaluate the performance of the beamline by analysing experimental data collected during its initial commissioning.
Measurements of the \pbar{} and \positron{} beam parameters were primarily made using instruments inserted into the beamline at the locations labelled with bold text in \mbox{Figure \ref{fig:alpha-schematic}}.
Diagnostic devices \cite{ALPHA-2014-1} installed at the `AT stick' and `CT stick' (see \mbox{Fig. \ref{fig:alpha-schematic}}) were used to measure the beam parameters within region one, while measurements along the new beamline were primarily made at the BDS, PDS and LDS.
Phosphor-backed micro channel plate (MCPs) and cameras mounted at these locations were used to destructively image the transverse profiles of \pbar{} and \positron{} bunches in transit along the beamline \cite{ALPHA-2009}.
\par
In addition, each diagnostics station is equipped with a Faraday Cup (FC) for measuring the total charge deposited by an incident electron or positron bunch.
FC measurements were not used to characterise \pbar{} bunches, as the total charge in each bunch is below the sensitivity of our readout electronics.
\par
As well as in-vacuum diagnostic devices, the ALPHA apparatus includes a wide range of external annihilation detectors.
Beam losses during particle transfers are monitored using an array of caesium iodide (CsI) scintillator crystals backed with silicon photomultiplier (SiPM) chips, which are mounted between each section of the beamline.
The CsI detectors are easily saturated by intense bursts of radiation with a long recovery time of \mbox{$\sim$ 700 $\mu$s}, and are unable to resolve the time structure of the annihilation signal from an entire \pbar{} or \positron{} bunch.
\par
Where improved time resolution or high-current capabilities are required, larger plastic scintillator panels backed with photomultiplier tubes (PMTs) or fast SiPM chips can be deployed around key areas of the experiment.
Since these detectors are read out at a much higher time resolution ($\lesssim$ 2 ns) than the length of a typical \pbar{} or \positron{} pulse \mbox{($\sim$ 1 $\mu$s)}, they can be used for destructive, single-shot measurements of the bunch time structure at a given point along the beamline.

\subsection{Initial Setup}
Installation of the new beamline was completed at CERN between May -- July 2018, in parallel with the construction of the \alphag{} experiment.
Commissioning was completed by November 2018, with \pbar{} and \positron{} bunches successfully transferred to both of the ALPHA \antihydrogen{} synthesis traps.
\par
The beamline was initially configured to transfer \positron{} bunches into the \alphatwo{} experiment along a horizontal beam path.
The magnetic alignment of the beamline was corrected by adjusting the orientations of the beamline module end coils (AGBL02).
After each adjustment, the positron beam position was measured using an MCP mounted at the AT stick (see \mbox{Fig. \ref{fig:alpha-schematic}}).
Upon completion of the preliminary alignment process, both \pbar{} and \positron{} bunches could be imaged at regular intervals along the full length of the horizontal beamline.
\par
No further alignment was necessary to transfer positrons into the \alphatwo{} experiment.
Transfers along the horizontal length of the beamline were initially tested by sending \positron{} bunches directly through the \alphatwo{} Penning trap.
Up to \mbox{$\left( 10.1 \pm 0.7 \right) \times 10^{6}$} \positron{} per shot were transferred without detecting significant annihilation losses along the beamline.
\begin{figure}
\hspace{-18pt}
\centering{
\includegraphics[scale=0.37, trim=15 15 13 15, clip]{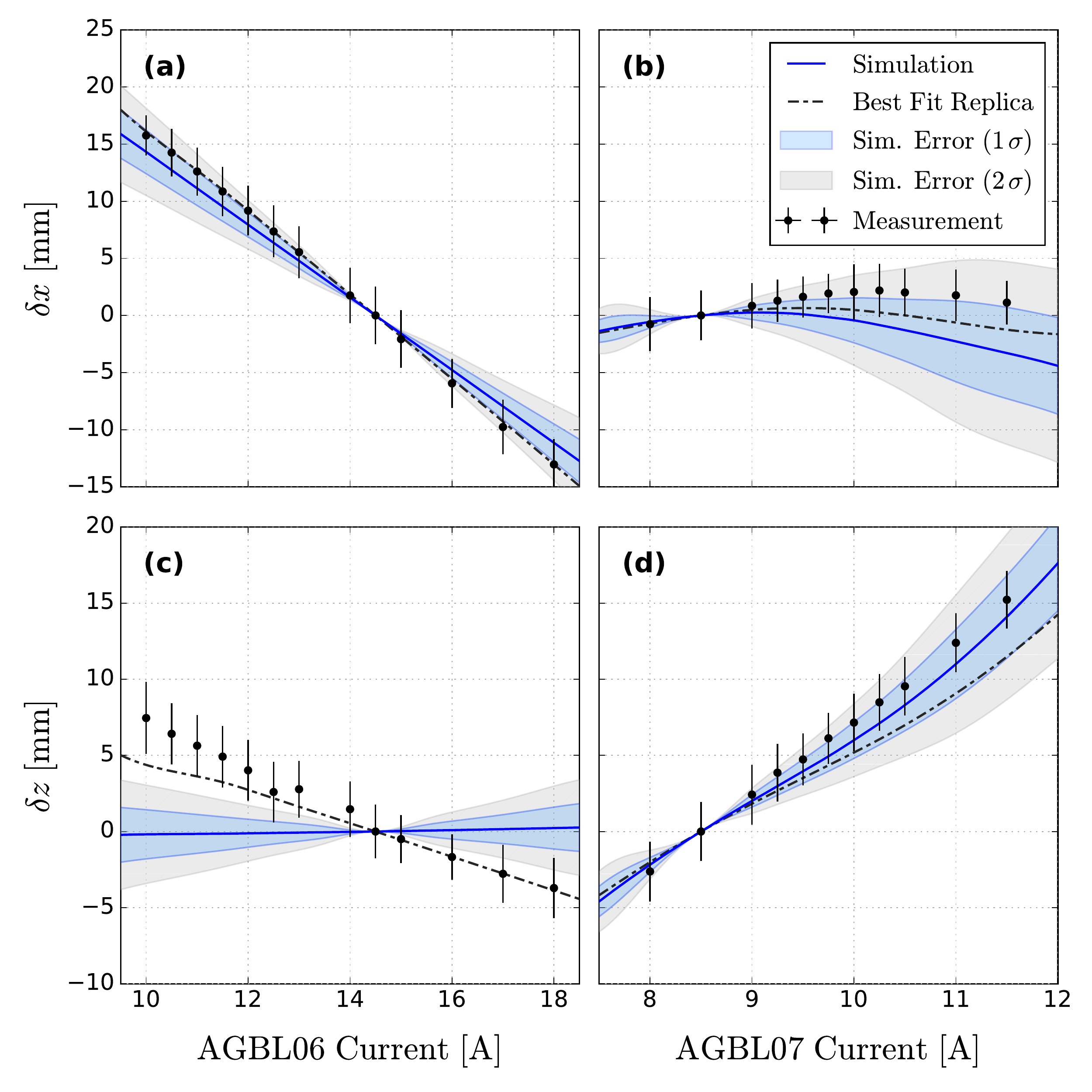}
}
\caption{Antiproton beam position at the LDS as a function of the AGBL06 and AGBL07 interconnect magnet currents. The beam position has been separated into components along the $x$ and $z$ axes of \mbox{Fig. \ref{fig:interconnect}(a)}. The experimental data were collected while the \alphag{} external solenoid was not energised. An equivalent, simulated steering curve for each data set is shown as a solid red line. The shaded intervals indicate the uncertainty in the simulated beam position due to mechanical errors along the full \pbar{} beamline.}
\label{fig:steering-data}
\end{figure}
\subsection{Beam Steering to \alphag{}}
Initial commissioning of the interconnect magnets was carried out by transferring \pbar{} and \positron{} bunches around a sharp right-angled turn and imaging them using an MCP mounted at the LDS.
Initial setpoints for the interconnect magnets (\mbox{Table \ref{table:magnets}}) were chosen based on the results of numerical particle tracing simulations.
In each case, only small corrections to these setpoints were required to locate the beam on the MCP.
\mbox{Figure \ref{fig:lds-mcp-images}} shows the transverse beam profiles of \mbox{50 eV} \pbar{} and \positron{} bunches that were imaged at the LDS.
In both images, the beam position has been chosen to maximise the number of particles delivered to the \alphag{} experiment, and is not necessarily aligned to the centre of the MCP.
By repeatedly imaging \pbar{} bunches extracted to the LDS, the beam position was measured to be stable within \mbox{$\lesssim$ 0.06 mm} (one standard deviation), slightly better than the expected performance \mbox{(0.15 mm)} based on the specified stability of the beamline magnet power supplies (\mbox{60 mA}).
\par
Integrated along the major and minor axes of their elliptical beam envelopes, the \pbar{} and \positron{} bunches shown in \mbox{Fig. \ref{fig:lds-mcp-images}} are normally distributed about their respective centres.
The \pbar{} beam profile in \mbox{Fig. \ref{fig:lds-mcp-images}(a)} has a mean radius of \mbox{3.2 mm} with the MCP immersed in a magnetic field of \mbox{$\sim$ 270 Gauss}.
This corresponds to a beam envelope of \mbox{0.53 mm} in the stronger \mbox{1.0 T} magnetic field of the \alphag{} experiment, consistent with the calculations shown in \mbox{Fig. \ref{fig:numerical-dynamics}(c)}.
The positron beam profile shown in \mbox{Fig. \ref{fig:lds-mcp-images}}(d) is more elliptical, with a radius of \mbox{1.96 mm} along its minor axis and \mbox{4.56 mm} along its major axis.
This is consistent with numerical simulations of the beamline, which predict that this elongation is caused by mixing of the beam's transverse and longitudinal degrees of freedom.
\par
The trajectories of charged particles through the interconnect were investigated by scanning the currents in select magnets.
\mbox{Figure \ref{fig:steering-data}} shows the \pbar{} beam position measured at the LDS as a function of the AGBL06 and AGBL07 magnet currents.
Since a limited amount of time was available for characterisation of the beamline, these data were collected while the \alphag{} external solenoid was not energised.
To compare these measurements against the expected performance of the interconnect in this scenario, a set of particle tracing simulations was carried out (shown as solid lines in \mbox{Fig. \ref{fig:steering-data}}).
\par
\begin{figure}[h!]
\centering{
\includegraphics[scale=0.44, trim=15 15 12 10, clip]{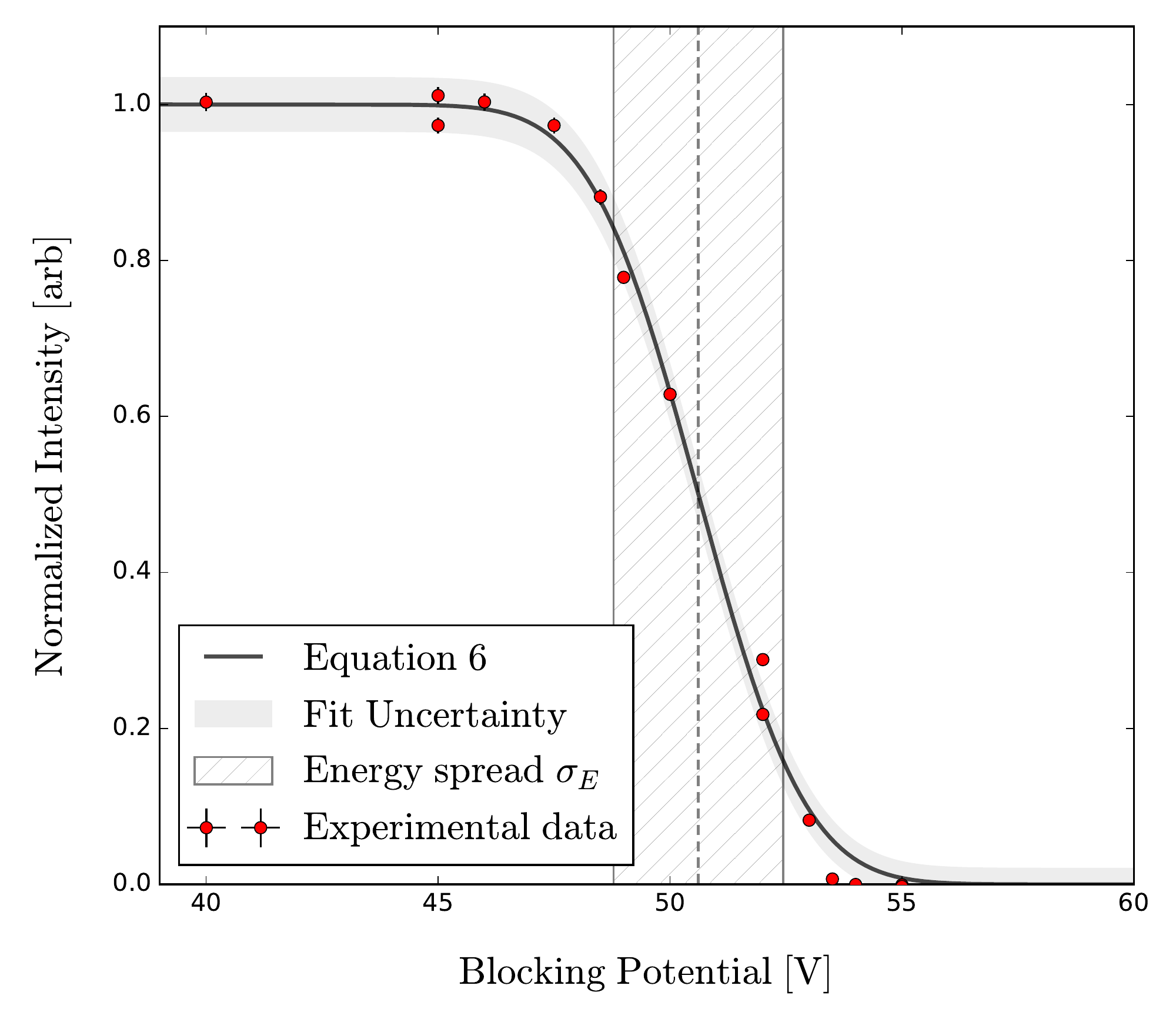}
}
\caption{Normalised \pbar{} beam intensity at the BDS as a function of the blocking voltage applied inside the \alphatwo{} Penning trap. \mbox{Equation \ref{eq:beam-blocking}} has been fitted to the experimental data to extract the beam energy and \pbar{} energy spread. Many of the error bars are too small to be visible.}
\label{fig:beam-blocking}
\end{figure}
There is reasonable agreement between the measured and simulated steering curves.
However, the AGBL06 magnets translate the beam along the $z$ axis of \mbox{Fig. \ref{fig:interconnect}(a)} in a way that is not predicted by the simulations.
The shaded bands in \mbox{Fig. \ref{fig:steering-data}} show the uncertainty in the simulated beam position due to typical mechanical errors along the full \pbar{} beamline.
As described in \mbox{Sec. \ref{sec:design}}, we estimate this uncertainty as the standard deviation of 200 replica simulations, each with a unique set of mechanical errors.
\par
To find the combination of errors that best reproduces our measurements, we evaluated the $\chi^{2}$ parameter for each set of replica simulations with the experimental data.
We show the specific replica with the best $\chi^{2}$ from our survey, and note that this replica broadly reproduces the trends seen in the experimental data.
The results of our survey suggest that the trends in \mbox{Fig. \ref{fig:steering-data}} arise due to a combination of many small errors distributed along the beamline, rather than a strong localised imperfection.
However, the system proved adaptable enough that good steering was achieved irrespective of these errors.
\par
The efficiency of positron transfers through the interconnect was estimated by comparing FC measurements taken at the PDS and LDS.
For bunches of up to \mbox{$3.8 \times 10^{7}$} \positron{}, we find that \mbox{$\left( 93 \pm 3 \right) \%$} of extracted positrons arrive at the LDS.
Estimating the \pbar{} transfer efficiency is more challenging, as our FC readout electronics are not sensitive enough to detect fewer than $10^{5}$ antiprotons.
However, no significant losses were observed using the CsI detectors spaced along the beamline, implying that any transfer losses are small.

\subsection{Bunch Structure}
As shown in \mbox{Sec. \ref{sec:beam-capture}}, the number of particles captured after each transfer is strongly dependent on the bunch length delivered to the \alphag{} Penning trap.
During commissioning of the new beamline, we tuned the extraction and capture of \pbar{} and \positron{} bunches to maximise the number of particles that were captured.
Extensive measurements of the beam energy distributions and bunch lengths were used to inform this effort.
\par
The antiproton energy distribution was measured by repeatedly imaging \pbar{} bunches at the BDS, while applying blocking voltages to electrodes inside the \alphatwo{} Penning trap.
\mbox{Figure \ref{fig:beam-blocking}} shows the beam intensity at the BDS as a function of the on-axis blocking potential.
At the start of each trial, a fixed fraction of antiprotons were released from the CT by manipulating the electric potential along the Penning trap axis.
The resulting annihilations were counted using a pair of PMT-backed scintillator panels, and used to normalise the beam intensity against shot-to-shot variations in the initial number of \pbar{}.
\par
Assuming a Gaussian distribution of parallel energies, the beam intensity is modelled by the expression \\
\begin{equation} \label{eq:beam-blocking}
f \left( \phi \right) = \frac{N}{2} \left[ 1 - \mathrm{erf} \left( \frac{q \phi - \langle E_{\parallel} \rangle}{\sqrt{2} \sigma_{E}} \right) \right] \,,
\end{equation} \\
where $N$ is the number of particles per bunch, $\phi$ is the on-axis blocking potential, and $\langle E_{\parallel} \rangle$ and $\sigma_{E}$ are the centroid beam energy and energy spread, respectively.
By fitting \mbox{Eq. \ref{eq:beam-blocking}} to the data in \mbox{Fig. \ref{fig:beam-blocking}}, we find that each \pbar{} bunch has a mean energy of \mbox{$\left( 50.6 \pm 0.1 \right)$ eV} and an energy spread of \mbox{$\left( 1.8 \pm 0.1 \right)$ eV}.
The potentials used to extract \pbar{} bunches from the CT, shown in \mbox{Fig. \ref{fig:catching-trap}(b)}, were optimised to reduce the energy spread.
\par
The extraction of \positron{} bunches from the PA was also optimised, producing an energy spread of \mbox{$\left(2.9 \pm 0.6 \right)$ eV} around a mean energy of \mbox{$\left( 48.1 \pm 0.4 \right)$ eV}.
These wide energy distributions result in elongation of the \pbar{} and \positron{} bunches, ultimately resulting in particles losses upon re-capture in the ALPHA Penning traps.
We anticipate that further optimisation of the positron and antiproton energy distributions will be possible after detailed studies of the beam extraction process \cite{Natisin-2016-2}.
\par
\begin{figure}
\centering{
\includegraphics[scale=0.42]{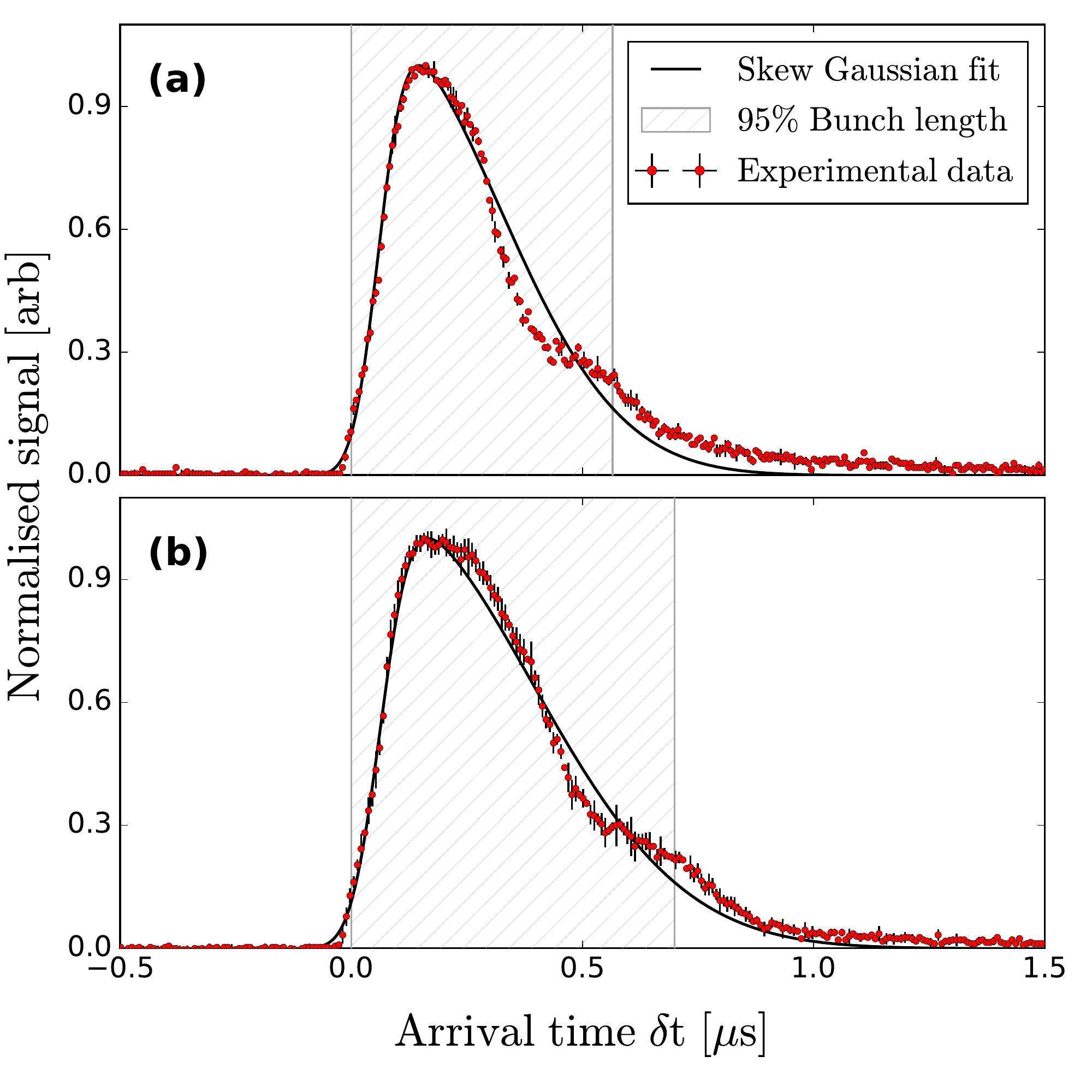}
}
\caption{Measurement showing the time structure of a \mbox{48 eV} \positron{} bunch (a) immediately before the interconnect and (b) at the LDS. Data was collected using a PMT-backed scintillator panel installed beside the interconnect, and averaged over three identical trials. The error bars reflect the standard error of the distribution of measurements. To aid comparison, the horizontal axis is shifted relative to the onset of each signal.}
\label{fig:bunch-structure}
\end{figure}

In addition, direct measurements of the \pbar{} and \positron{} bunch lengths were taken at several points along the beamline.
Plastic scintillator panels backed with PMTs were used to record the time structure of the annihilations when a physical barrier was placed into the beam path.
\mbox{Figure \ref{fig:bunch-structure}}(a) shows the normalised PMT signal that was recorded when a \mbox{48 eV} \positron{} bunch was made to annihilate immediately before entering the interconnect.
\par
The experimental data has been averaged over three identical trials, and the horizontal axis shifted relative to the onset of the annihilation signal.
A smooth curve has been empirically fitted to the data to extract the \mbox{95 \%} bunch length \mbox{$\sigma_{\tau}$ = 0.57 $\mu$s}.
\par
\mbox{Figure \ref{fig:bunch-structure}(b)} shows a similar measurement for a \mbox{48 eV} \positron{} bunch annihilating at the LDS, approximately \mbox{75 cm} further along the beam path.
After passing through the interconnect, the bunch length has increased by \mbox{0.13 $\mu$s} to \mbox{$\sigma_{\tau}$ = 0.70 $\mu$s}.
This increase may be driven by \positron{} bunches having strongly divergent longitudinal phase spaces, and also being subject to significant space charge forces while in transit.
A detailed investigation of factors that contribute to this increase in bunch length is beyond the scope of this paper.
\par
Similar measurements were carried out along the beamline for antiproton bunches with a range of energies up to \mbox{75 eV}.
The nominal \pbar{} beam energy was chosen to minimise the antiproton bunch length in the vicinity of the \alphag{} experiment.
Antiprotons extracted to the LDS at an energy of \mbox{$\sim$ 50 eV} have a \mbox{95 \%} bunch length of \mbox{$\left( 1.66 \pm 0.02 \right)$ $\mu$s}.
\begin{figure}[h]
\renewcommand{\abovecaptionskip}{-5pt}
\renewcommand{\belowcaptionskip}{-5pt}
\centering{
\includegraphics[scale=0.44, trim=12 15 0 0, clip]{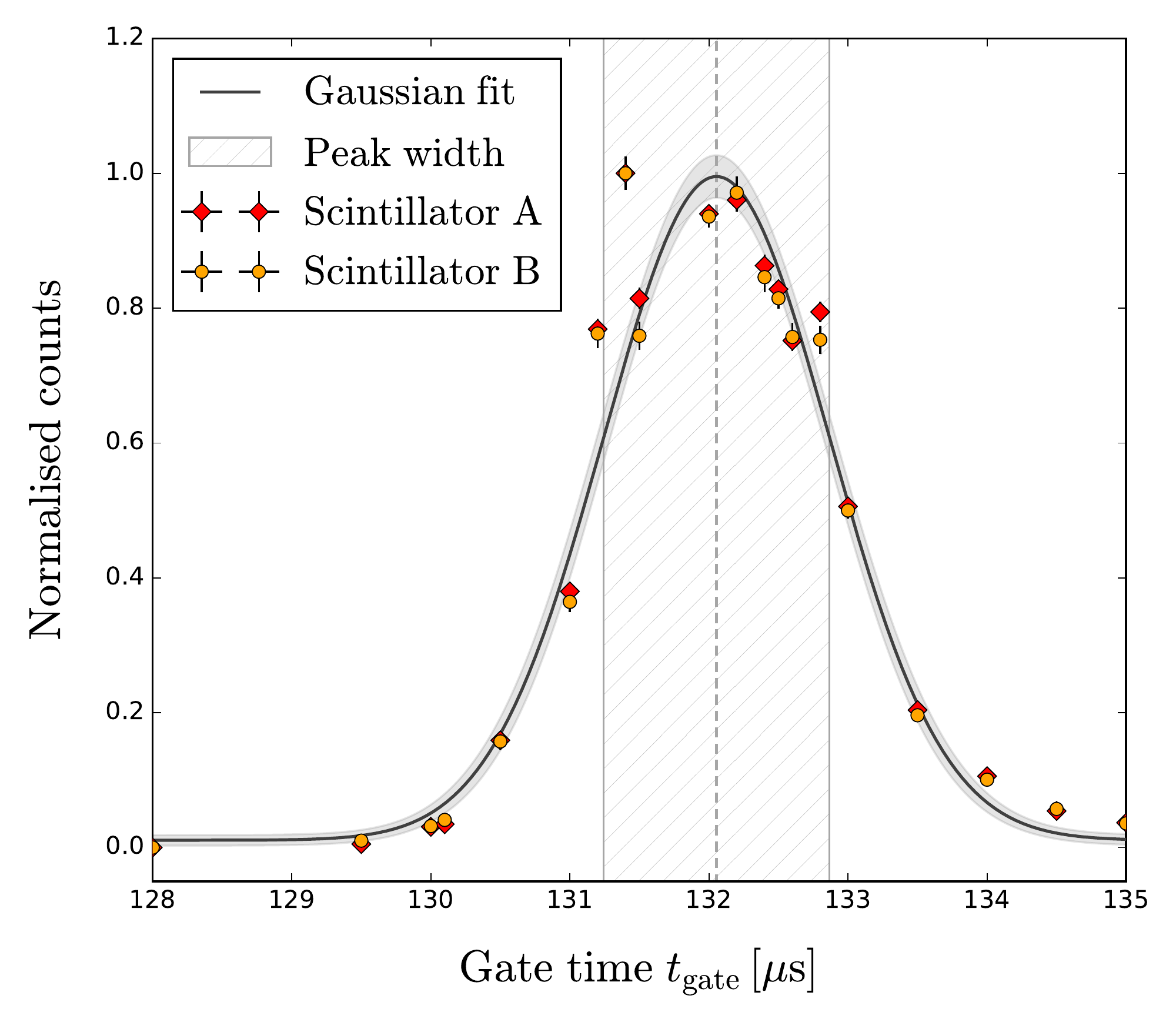}
}
\caption{Number of annihilations during a five second window after \pbar{} bunches are caught inside the \alphag{} Penning trap, for a range of gate times between \mbox{128 $\mu$s} and \mbox{135 $\mu$s}. The number of annihilations has been normalised against shot-to-shot variation in the initial number of \pbar{}. The error bars reflect Poisson counting uncertainties.}
\label{fig:gate-scan-data}
\end{figure}

\subsection{Bunch Capture}
As discussed in \mbox{Sec. \ref{sec:beam-capture}}, \pbar{} and \positron{} bunches delivered by the beamline must ultimately be captured in one of the two \antihydrogen{} synthesis traps.
During the initial setup of the beamline, bunches of up to $\sim 10^{7}$ \positron{} were steered directly through \alphatwo{} without being captured.
However, the number of particles that could be captured from each bunch was limited by the long bunch length delivered to the Penning trap.
After optimisation of the potentials used to catch \positron{} bunches, we estimate that \mbox{$\left( 71 \pm 5 \right)\%$} of positrons delivered to the \alphatwo{} experiment were captured per shot.
This efficiency was sufficient for \antihydrogen{} production at a rate consistent with the performance of the experiment during 2017, when the PA was directly connected to the \alphatwo{} apparatus \cite{ALPHA-2017-2}.
\par
Separately, \pbar{} bunches were extracted from the CT and captured in the \alphag{} Penning trap.
The antiproton gate time was optimised by attempting to capture \pbar{} bunches in a \mbox{105 mm} long, \mbox{140 V} electrostatic potential well, and counting the number of annihilations during a \mbox{5 second} window after the antiprotons were expected to arrive.
Without electron cooling, the captured antiprotons escape from confinement or annihilate with background gases after less than \mbox{$\sim$ 0.8 s} inside the trap volume.
\mbox{Figure \ref{fig:gate-scan-data}} shows the number of annihilations counted by a pair of SiPM-backed scintillator panels as a function of the \pbar{} gate time $t_\mathrm{gate}$.
As in \mbox{Fig. \ref{fig:beam-blocking}}, the number of annihilations has been normalised against shot-to-shot variations in the initial number of \pbar{}.
The resulting curve represents an experimental equivalent to the simulations shown in Fig. \ref{fig:gate-scan-sim}.
\par
A Gaussian function has been fitted to the data in \mbox{Fig. \ref{fig:gate-scan-data}} to extract the optimal gate time \mbox{(132 $\mu$s)} and the width of the measured curve \mbox{($0.81 \pm 0.04$ $\mu$s)}.
By comparing the width of the fitted curve to the simulations shown in \mbox{Fig. \ref{fig:gate-scan-sim}} (described in \mbox{Sec. \ref{sec:beam-capture}}), we estimate the \pbar{} bunch length within the Penning trap to be \mbox{$\left( 2.4 \pm 0.2 \right)$ $\mu$s}.
For this bunch length, numerical simulations predict that \mbox{$\left( 78 \pm 3 \right)$ \%} of antiprotons are captured from each bunch, excluding other loss mechanisms that may occur outside of the Penning trap.
\par
After optimising the gate time, electron cooling was demonstrated by loading around \mbox{$2.5 \times 10^{7}$} \electron{} into the Penning trap before the arrival of each \pbar{} bunch.
Electrons were loaded into the trap using an electron gun mounted at the LDS, and expanded radially \cite{Huang-1997} before each \pbar{} transfer.
After establishing electron cooling, antiprotons were held inside the \alphag{} Penning trap for more than \mbox{300 seconds} without any significant loss of particles.
\par
Positron bunches were also captured inside the \alphag{} experiment and confined for long timescales.
Due to their long bunch length, we expect to lose large numbers of positrons inside the Penning trap at the end of each transfer.
These losses were minimised by tuning the electric potentials used to capture positrons inside the \alphag{} experiment.
After optimisation of the trapping potentials, up to \mbox{$\sim 10^{7}$ \positron{}} could be captured from each bunch delivered to the \alphag{} Penning trap.
This number of positrons is more than sufficient for \antihydrogen{} production using the same techniques that have already been demonstrated using the \alphatwo{} experiment \cite{ALPHA-2017-2}.

\section{SUMMARY AND OUTLOOK} \label{sec:summary}
The ALPHA collaboration has recently implemented a novel multi-species beamline for low energy \pbar{} and \positron{} beams extracted from Penning-Malmberg traps.
A combination of semi-analytical and numerical methods were used during the development of the beamline, allowing insight into the relevant physics while minimising the overall computational effort.
Measurements have shown that large numbers of positrons and antiprotons can already be transported efficiently throughout the ALPHA apparatus.
The successful operation and achieved performance of this beamline were critical to the initial commissioning of the \alphag{} experiment, and several physics measurements made with the \alphatwo{} experiment during 2018 \cite{ALPHA-2021-1}.
\par
This work will play an important role in future physics measurements at ALPHA, such as precision \antihydrogen{} spectroscopy and direct measurements of antimatter's gravitational acceleration.
The techniques presented here are also expected to be useful for positron and ion trap experiments where low-energy beams are derived from Penning traps with strong axial magnetic fields.
\par
In the future, the performance of the beamline will be improved by optimising the longitudinal dynamics of \pbar{} and \positron{} bunches delivered to the antihydrogen synthesis traps.
Shorter bunches may be obtained with different extraction protocols, optimised by detailed computational and experimental studies of the methods used to extract beams from the CT and PA \cite{Natisin-2016-2}.
There are ongoing experimental efforts at ALPHA to compress the temporal profiles of \pbar{} and \positron{} bunches in-flight, using time-dependent electric potentials applied along parts of the beamline \cite{Aghion-2015}.

\section*{Acknowledgements}
This work was supported by: the European Research Council through its Advanced Grant programme (JSH); CNPq, FAPERJ, RENAFAE (Brazil); NSERC, CFI, NRC/TRIUMF, EHPDS/EHDRS (Canada); FNU (Nice Centre), Carlsberg Foundation (Denmark); ISF (Israel); STFC, EPSRC, the Royal Society and the Leverhulme Trust (UK); DOE, NSF (USA); and VR (Sweden). 
We are grateful for the efforts of the CERN AD team, without which these experiments could not have taken place. 
We thank Jacky Tonoli (CERN). 
We thank D. Tommasini and A Milanese (CERN) for the fabrication of Helmholtz coils for the diagnostic stations, and L. Clarke, M. Booth, A. Nichols, and J. Boehm (STFC, UK) for project management and manufacturing of the beamline magnets, vacuum chambers and support structures.

\bibliographystyle{apsrev4-2}
\bibliography{Johnson_ALPHA_beamline_16Nov2022}

\end{document}